\documentclass[a4,11pt]{article}
\usepackage{graphicx} 
\usepackage{amsmath}
\usepackage{amsfonts}
\usepackage{latexsym}
\usepackage{amssymb}
\makeatletter
 
  \@addtoreset{equation}{section}
 \makeatother

\begin{document}
\title{Analytical Approximation for Non-linear FBSDEs \\
with Perturbation Scheme~\footnote{
This research is supported by CARF (Center for Advanced Research in Finance) and 
the global COE program ``The research and training center for new development in mathematics.''
All the contents expressed in this research are solely those of the authors and do not represent any views or 
opinions of any institutions. 
The authors are not responsible or liable in any manner for any losses and/or damages caused by the use of any contents in this research.
}}

\author{Masaaki Fujii\footnote{Graduate School of Economics, The University of Tokyo},
Akihiko Takahashi\footnote{Graduate School of Economics, The University of Tokyo}
}
\date{
First version: June 1, 2011\\ 
This version: January 20, 2012
}
\maketitle



\newtheorem{definition}{Definition}
\newtheorem{assumption}{$[$ A}
\newtheorem{condition}{$[$ C}
\newtheorem{lemma}{Lemma}
\newtheorem{proposition}{Proposition}
\newtheorem{theorem}{Theorem}
\newtheorem{remark}{Remark}
\newtheorem{example}{Example}
\newtheorem{corollary}{Corollary}
\def\n{{\bf n}}
\def\A{{\bf A}}
\def\B{{\bf B}}
\def\C{{\bf C}}
\def\D{{\bf D}}
\def\E{{\bf E}}
\def\F{{\bf F}}
\def\G{{\bf G}}
\def\H{{\bf H}}
\def\I{{\bf I}}
\def\J{{\bf J}}
\def\K{{\bf K}}
\def\L{{\bf L}}
\def\M{{\bf M}}
\def\N{{\bf N}}
\def\O{{\bf O}}
\def\P{{\bf P}}
\def\Q{{\bf Q}}
\def\R{{\bf R}}
\def\S{{\bf S}}
\def\T{{\bf T}}
\def\U{{\bf U}}
\def\V{{\bf V}}
\def\W{{\bf W}}
\def\X{{\bf X}}
\def\Y{{\bf Y}}
\def\Z{{\bf Z}}
\def\cala{{\cal A}}
\def\calb{{\cal B}}
\def\calc{{\cal C}}
\def\cald{{\cal D}}
\def\cale{{\cal E}}
\def\calf{{\cal F}}
\def\calg{{\cal G}}
\def\calh{{\cal H}}
\def\cali{{\cal I}}
\def\calj{{\cal J}}
\def\calk{{\cal K}}
\def\call{{\cal L}}
\def\calm{{\cal M}}
\def\caln{{\cal N}}
\def\calo{{\cal O}}
\def\calp{{\cal P}}
\def\calq{{\cal Q}}
\def\calr{{\cal R}}
\def\cals{{\cal S}}
\def\calt{{\cal T}}
\def\calu{{\cal U}}
\def\calv{{\cal V}}
\def\calw{{\cal W}}
\def\calx{{\cal X}}
\def\caly{{\cal Y}}
\def\calz{{\cal Z}}
%
\def\sskip{\hspace{0.5cm}}
\def\simleq{ \raisebox{-.7ex}{\em $\stackrel{{\textstyle <}}{\sim}$} }
\def\leqsim{ \raisebox{-.7ex}{\em $\stackrel{{\textstyle <}}{\sim}$} }
\def\ep{\epsilon}
\def\half{\frac{1}{2}}
\def\iku{\rightarrow}
\def\Iku{\Rightarrow}
\def\ikup{\rightarrow^{p}}
\def\inclusion{\hookrightarrow}
\def\cadlag{c\`adl\`ag\ }
\def\up{\uparrow}
\def\down{\downarrow}
\def\doti{\Leftrightarrow}
\def\douti{\Leftrightarrow}
\def\dochi{\Leftrightarrow}
\def\douchi{\Leftrightarrow}%
\def\yy{\\ && \nonumber \\}
\def\y{\vspace*{3mm}\\}
\def\nn{\nonumber}
\def\be{\begin{equation}}
\def\ee{\end{equation}}
\def\bea{\begin{eqnarray}}
\def\eea{\end{eqnarray}}
\def\beas{\begin{eqnarray*}}
\def\eeas{\end{eqnarray*}}
%
\def\hd{\hat{D}}
\def\hv{\hat{V}}
\def\hsd{{\hat{d}}}
\def\hx{\hat{X}}
\def\hsx{\hat{x}}
\def\bsx{\bar{x}}
\def\bsd{{\bar{d}}}
\def\bx{\bar{X}}
\def\ba{\bar{A}}
\def\bb{\bar{B}}
\def\bc{\bar{C}}
\def\bv{\bar{V}}
\def\balpha{\bar{\alpha}}
\def\bbalpha{\bar{\bar{\alpha}}}
\def\combi{\l(\begin{array}{c}\alpha\\ \beta \end{array}\r)}
\def\f{^{(1)}}
\def\s{^{(2)}}
\def\ss{^{(2)*}}
\def\l{\left}
\def\r{\right}
\def\a{\alpha}
\def\b{\beta}
\def\L{\Lambda}

\def\E{{\bf E}}
\def\P{{\bf P}}
\def\Q{{\bf Q}}
\def\R{{\bf R}}

\def\calf{{\cal F}}
\def\calp{{\cal P}}
\def\calq{{\cal Q}}

\def\ep{\epsilon}
\def\del{\delta}
\def\al{\alpha}
\def\part{\partial}
\def\ol{\overline}

\def\yy{\\ && \nonumber \\}
\def\y{\vspace*{3mm}\\}
\def\nn{\nonumber}
\def\be{\begin{equation}}
\def\ee{\end{equation}}
\def\bea{\begin{eqnarray}}
\def\eea{\end{eqnarray}}
\def\beas{\begin{eqnarray*}}
\def\eeas{\end{eqnarray*}}
\def\l{\left}
\def\r{\right}
\vspace{10mm}

\begin{abstract}
In this work, we have presented a simple analytical approximation scheme for 
generic non-linear FBSDEs. By treating the interested system as the 
linear decoupled FBSDE perturbed with non-linear generator and feedback terms,
we have shown that it is possible to carry out a recursive approximation 
to an arbitrarily higher order, where the required calculations in each order 
are equivalent to those for standard European contingent claims.
We have also applied the perturbative method to the PDE framework following 
the so-called Four Step Scheme. The method is found to 
render the original non-linear PDE into a series of standard parabolic linear PDEs.
Due to the equivalence of the two approaches, it is also possible to 
derive approximate analytic solution for the non-linear PDE by 
applying the asymptotic expansion to the corresponding probabilistic model.
Two simple examples are provided to demonstrate how the perturbation works
and show its accuracy relative to known numerical techniques.
The method presented in this paper may be useful for various 
important problems which have eluded analytical treatment so far.

\end{abstract}
\vspace{17mm}

\newpage
\section{Introduction}
In this paper, we propose a simple analytical approximation for
backward stochastic differential equations (BSDEs).
These equations were introduced by Bismut (1973)~\cite{Bismut}
for the linear case and later by Pardoux and Peng (1990)~\cite{Peng1}
for the general case,  and have earned strong academic interests since then.
They are particularly relevant for the pricing of contingent claims
in constrained or incomplete markets, and for the study 
of recursive utilities as presented by
Duffie and Epstein (1992)~\cite{Duffie_Epstein}.
For a recent comprehensive study with financial applications, one may consult
Yong and Zhou (1999)~\cite{zhou}, Ma and Yong (2000)~\cite{Ma} and references therein.

The importance of BSDEs, or more specifically non-linear FBSDEs
which have non-linear generators coupled with some state processes
satisfying the forward SDEs, has risen greatly in recent years also among practitioners.
The collapse of major financial institutions followed by the drastic reform of regulations
make them well aware of the importance of counterparty risk management,
credit value adjustments (CVA) in particular.
Even in a very simple setup, if there exists asymmetry in the credit risk 
between the two parties, the relevant dynamics of portfolio value 
follows a non-linear FBSDE as clearly shown by Duffie and Huang (1996)~\cite{Duffie}.
We have recently found that the asymmetric treatment of collateral 
between the two parties also leads to a non-linear FBSDE~\cite{asymmetric_collateral}. 
Furthermore, in May 2010, regulators were forced to realize the importance of mutual interactions
and feedback loops in the trading activities among financial firms, shocked by the astonishing
flash crash of the Dow Jones index by almost $1,000$ points. Once we take the
feedback effects from the behavior of major players into account, we naturally 
end up with complicated coupled FBSDEs.

Unfortunately, however, an explicit solution for a FBSDE is only known for
a simple linear example. In the last decade, several techniques have been introduced by
researchers, but they tend to be quite complicated for practical applications.
They either require one to solve non-linear PDEs, which are very difficult in general, 
or resort to quite time-consuming simulation.
Although regression based Monte Carlo simulation has been rather popular among 
practitioners for the pricing of callable products, the appropriate choice 
of regressors and attaining numerical stability becomes a more subtle issue for a general FBSDE. 
In fact, in clear contrast to the pricing of callable products,
one cannot tell if the price goes up or down when one improves the 
regressors, which makes it particularly difficult to select the appropriate basis functions.

In this paper, we present a simple analytical approximation scheme 
for the non-linear FBSDEs coupled with generic Markovian state processes.
We have perturbatively expanded the non-linear terms around the linearized FBSDE,
where the expansion can be made recursively to an arbitrary higher order.
In each order of approximation, the required calculations are 
equivalent to those for standard European contingent claims.
In order to carry out the perturbation scheme,
we need to express the backward components explicitly in terms of the forward components for
each order of approximation.
For that purpose, we propose to use the asymptotic expansion of volatility for the forward components,
which is now widely adopted to price various European contingent claims and compute optimal portfolios
(See, for examples \cite{KT,STT,T,asymptotic3,TY}
and references therein for the recent developments and review.).
In the case when the underlying processes have known distributions, of course,  we can directly 
proceed to a higher order approximation without resorting to an asymptotic expansion.

We have also studied a perturbation scheme in the PDE framework, or in the so-called Four Step Scheme~\cite{fourstep},
for the generic fully-coupled non-linear FBSDEs.
We have shown that our perturbation method renders the original non-linear PDE into the series of classical
linear parabolic PDEs, which are easy to handle with standard techniques. 
We then provided the corresponding probabilistic framework by using the 
equivalence between the two approaches. 
We have shown that, also in this case, the required calculations in a given order
are  equivalent to those for the classical  European contingent claims. 
As a by-product, by applying the asymptotic expansion method to the 
corresponding probabilistic model, it was actually found possible to derive an
{\it analytic expression} for the solution of the non-linear PDE 
up to a given order of perturbation.
Therefore, our method can be interpreted as a practical implementation of the Four Step Scheme
in the perturbative approach.

The organization of of the paper is as follows:
In Section 2, we will explain our new approximation scheme 
with perturbative expansion for generic decoupled FBSDEs. Then, in Section 3, we shall apply it 
to the two concrete examples to demonstrate how it works and test its 
numerical performance. One of them allows
a direct numerical treatment by a simple PDE and hence it is easy to compare the two methods.
In the second example, we will consider a slightly more complicated model.
We compare our approximation result to 
the detailed numerical study recently carried out by Gobet et al. (2005)~\cite{Gobet} 
using a regression-based Monte Carlo simulation. 
In Section 4, we explain how to use standard asymptotic expansion procedures
to express the backward components explicitly when the forward components
do not have known distributions.
In Section 5, we will give an extension of our method to the fully coupled non-linear FBSDEs under the 
PDE framework, and then formulate the equivalent probabilistic approach in Section 5.
Appendix contains slightly different scheme for coupled non-linear FBSDEs which may 
be useful for the actual application.

\section{Approximation Scheme}
\label{decoupled_approx}
\subsection{Setup}
Let us briefly describe the basic setup.
The probability space is taken as $(\Omega, \calf, P)$ and $T\in (0,\infty)$
denotes some fixed time horizon. $W_t=(W^1_t, \cdots, W^r_t)^*$, $0\leq t \leq T$ is
$\mathbb{R}^r$-valued Brownian motion defined on $(\Omega,\calf,P)$, and 
$(\calf_t)_{\{0\leq t\leq T\}}$ stands for $P$-augmented natural filtration 
generated by the Brownian motion.

We consider the following forward-backward stochastic differential equation (FBSDE)
\bea
dV_t&=&-f(X_t,V_t,Z_t)dt+Z_t\cdot dW_t \\
V_T&=&\Phi(X_T)
\eea
where $V$ takes the value in $\mathbb{R}$, and $X_t\in \mathbb{R}^d$ is assumed to follow a generic Markovian forward SDE 
\be
dX_t=\gamma_0(X_t)dt+\gamma(X_t)\cdot dW_t ~.
\ee
Here, we absorbed an explicit dependence on time to $X$ by allowing some of its components  
can be a time itself. $\Phi(X_T)$ denotes the terminal payoff where 
$\Phi(x)$ is a deterministic function of $x$.
The following approximation procedures can be applied in the same way 
also in the presence of coupon payments. 
$Z$ and $\gamma$ take values in $\mathbb{R}^r$ and $\mathbb{R}^{d\times r}$ respectively, 
and "$\cdot$" in front of the $dW_t$ represents the summation for the 
components of $r$-dimensional Brownian motion.
Throughout this paper, we are going to assume that the appropriate regularity conditions are satisfied 
for the necessary treatments.

\subsection{Perturbative Expansion for Non-linear Generator}
In order to solve the pair of $(V_t,Z_t)$ in terms of $X_t$, we extract the linear term from the 
generator $f$ and treat the residual non-linear term as the perturbation to the 
linear FBSDE. We introduce the perturbation parameter $\ep$, and then write 
the equation as
\bea
\label{mod_BSDE1}
dV_t^{(\ep)}&=&c(X_t)V_t^{(\ep)}dt-\ep g(X_t,V_t^{(\ep)},Z_t^{(\ep)})dt+Z_t^{(\ep)}\cdot dW_t  \\
V_T^{(\ep)}&=&\Phi(X_T) ~,
\label{mod_BSDE2}
\eea
where $\ep=1$ corresponds to the original model 
by~\footnote{Or, one can consider $\ep=1$ as simply a parameter convenient to 
count the approximation order. The actual quantity that should be small for the 
approximation is the residual part $g$.}
\be
f(X_t,V_t,Z_t)=-c(X_t)V_t+g(X_t,V_t,Z_t) ~.
\ee
Usually, $c(X_t)$ corresponds to the risk-free interest rate at time $t$, but it is not a necessary
condition. One should choose the linear term in such a way that the residual non-linear term
becomes as small as possible to achieve better convergence. A possible linear term  
$\theta(X)Z$ in the driver $f$  can be absorbed by the measure change and hence the simple reinterpretation of the 
drift term of the forward components $\gamma_0$ results in the form (\ref{mod_BSDE1}). See also the discussion in 
Appendix.

Now, we are going to expand the solution of BSDE (\ref{mod_BSDE1}) and (\ref{mod_BSDE2}) 
in terms of $\ep$: that is, suppose $V_t^{(\ep)}$ and $Z_t^{(\ep)}$ are expanded as
\bea
V_t^{(\ep)}&=& V_t^{(0)}+\ep V_t^{(1)}+\ep^2 V_t^{(2)}+\cdots\\
Z_t^{(\ep)}&=& Z_t^{(0)}+\ep Z_t^{(1)}+\ep^2 Z_t^{(2)}+\cdots. 
\eea

Once we obtain the solution up to the certain order, say $k$ for example, then by putting $\ep=1$,
\be
\tilde{V}_t= \sum_{i=0}^k  V_t^{(i)}, \hspace{10mm}
\tilde{Z}_t= \sum_{i=0}^k  Z_t^{(i)}
\ee
is expected to provide a reasonable approximation for the original model 
as long as the residual term is small enough to allow the perturbative treatment. 
As we will see, $V_t^{(i)}$ and $Z_t^{(i)}$, the corrections to each order can be calculated recursively 
using the results of the lower order approximations.

\subsection{Recursive Approximation for Perturbed linear FBSDE}
\subsubsection{Zero-th Order}\label{decoupled_zeroth}
For the zero-th order of $\ep$, one can easily see the following equation 
should be satisfied:
\bea
dV_t^{(0)}&=&c(X_t)V_t^{(0)}dt+Z_t^{(0)}\cdot dW_t \yy
V_T^{(0)}&=&\Phi(X_T) ~.
\eea
It can be integrated as
\be
V_t^{(0)}=E\left[\left.e^{-\int_t^T c(X_s)ds}\Phi(X_T)\right|\calf_t\right]
\label{V0formula}
\ee
which is equivalent to the pricing of a standard European contingent claim.

Since we have
\be
e^{-\int_0^T c(X_s)ds}\Phi(X_T)=V_0^{(0)}+\int_0^T e^{-\int_0^u c(X_s)ds}Z_u^{(0)}\cdot dW_u
\ee
it can be shown that, by applying Malliavin derivative $\cald_t$,  
\be
\cald_t\left(e^{-\int_0^T c(X_s)ds}\Phi(X_T)\right)=\int_t^T
\cald_t \left(e^{-\int_0^u c(X_s)ds}Z_u^{(0)}\right)\cdot dW_u
+e^{-\int_0^t c(X_s)ds}Z_t^{(0)}~.
\ee
Thus, by taking conditional expectation $E[\left.\cdot\right|\calf_t]$, we obtain
\be
Z_t^{(0)}=E\left[\left.\cald_t\left(e^{-\int_t^T c(X_s)ds}\Phi(X_T)\right)\right|\calf_t\right]~.
\ee

\subsubsection{First Order}
Now, let us consider the process $V^{(\ep)}-V^{(0)}$. One can see
that its dynamics is governed by 
\bea\label{1stV}
d\bigl(V_t^{(\ep)}-V_t^{(0)}\bigr)&=&c(X_t)\bigl(V_t^{(\ep)}-V_t^{(0)}\bigr)
-\ep g(X_t,V_t^{(\ep)},Z_t^{(\ep)})dt+\bigl(Z_t^{(\ep)}-Z_t^{(0)}\bigr)\cdot dW_t \nn \\
V_T^{(\ep)}-V_T^{(0)}&=&0~.
\eea
Now, by extracting the $\ep$-first order terms, we can once again recover the linear FBSDE
\bea
dV_t^{(1)}&=&c(X_t)V_t^{(1)}dt-g(X_t,V_t^{(0)},Z_t^{(0)})dt+Z_t^{(1)}\cdot dW_t \\
V_T^{(1)}&=&0~,
\eea
which leads to 
\be
V_t^{(1)}=E\left[\left.\int_t^T e^{-\int_t^u c(X_s)ds}g(X_u,V_u^{(0)},Z_u^{(0)})du\right|\calf_t\right]~
\ee
straightforwardly.
By the same arguments in the zero-th order example, we can express the 
volatility term as
\be
Z_t^{(1)}=E\left[\left.\cald_t\left(\int_t^T e^{-\int_t^u c(X_s)ds}g(X_u,V_u^{(0)},Z_u^{(0)})du\right)
\right|\calf_t\right]~.
\ee
From these results, we can see that the required calculation is nothing more difficult than
the zero-th order case as long as we have explicit expression for $V^{(0)}$ and $Z^{(0)}$.

\subsubsection{Second and Higher Order Corrections}
We can proceed the same way to the second order correction.
By extracting the $\ep$-second order terms from 
$V_t^{(\ep)}-(V_t^{(0)}+\ep V_t^{(1)})$, one can show  that
\bea
dV_t^{(2)}&=&c(X_t)V_t^{(2)}dt-\left(\frac{\part}{\part v}g(X_t,V_t^{(0)},Z_t^{(0)})V_t^{(1)}
+\nabla_z g(X_t,V_t^{(0)},Z_t^{(0)})\cdot Z_t^{(1)}\right)dt+Z_t^{(2)}\cdot dW_t \nn \\
V_T^{(2)}&=&0
\eea
is a relevant FBSDE, which is once again linear in $V_t^{(2)}$.
As before, it leads to the following expression straightforwardly:
\bea
V_t^{(2)}&=&E\left[\left.\int_t^T e^{-\int_t^u c(X_s)ds}\left(
\frac{\part}{\part v}g(X_u,V_u^{(0)},Z_u^{(0)})V_u^{(1)}+\nabla_z
g(X_u,V_u^{(0)},Z_u^{(0)})\cdot Z_u^{(1)}\right)du\right|\calf_t\right] \nn \\
\\
Z_t^{(2)}&=&E\left[\left.\cald_t\left(\int_t^T e^{-\int_t^u c(X_s)ds}\Bigl(
\frac{\part}{\part v}g(X_u,V_u^{(0)},Z_u^{(0)})V_u^{(1)}+\nabla_z
g(X_u,V_u^{(0)},Z_u^{(0)})\cdot Z_u^{(1)}\Bigr)du\right)\right|\calf_t\right]~.\nn\\
\eea
In the above calculation, we have assumed the driver function is differentiable.
If this is not the case, we need to approximate it using some smooth function or 
apply 
integration- by-parts technique for generalized Wiener functionals
(e.g. a composite functional of Dirac delta fucntion and 
a smooth Wiener functional).

In exactly the same way, one can derive an arbitrarily higher order correction.
Due to the $\ep$ in front of the non-linear term $g$, the system
remains to be linear in the every order of approximation.
However, in order to carry out explicit evaluation, we need to give
Malliavin derivative explicitly in terms of the forward components.
We will discuss this issue in the next.
\subsection{Evaluation of Malliavin Derivative}
Firstly, let us introduce a $d\times d$ matrix process $Y_{t,u}$, for $u\in[t,T]$,  
as the solution for the following forward SDE:
\bea
\label{YSDE1}
d(Y_{t,u})^i_j&=&\sum_{k=1}^d \Bigl(\part_k \gamma^i_0(X_u)(Y_{t,u})^k_j du+
\part_k \gamma^i(X_u)(Y_{t,u})^k_j \cdot dW_u\Bigr) \\
(Y_{t,t})^i_j&=&\del^i_j~,
\label{YSDE2}
\eea
where $\part_k$ denotes the differential with respect to the $k$-th component of $X$,
and $\del^i_j$ denotes Kronecker delta.

Now, for Malliavin derivative, we want to express, for $u\in[t,T]$,
\bea
E\left[\left.\cald_t
\left(e^{-\int_t^uc(X_s)ds}G(X_u)\right)\right|\calf_t\right]
\eea
in terms of $X_t$, where $G$ is a some deterministic function of $X$, in general.
Thank to the known chain rule of Malliavin derivative, we have
\bea
\cald_t\left(e^{-\int_t^uc(X_s)ds}G(X_u)\right)&=&\sum_{i=1}^d\Bigl\{e^{-\int_t^u c(X_s)ds}\part_i G(X_u) (\cald_t X^i_u) \nn \\
&&-e^{-\int_t^u c(X_s)ds}G(X_u)\left(\int_t^u \part_i c(X_s)(\cald_t X^i_s)ds\right)\Bigr\}.
\eea
Thus, it is enough for our purpose to evaluate  $(\cald_t X_u)$.
Since we have
\be
(\cald_t X^i_u)=\sum_{k=1}^d\Bigl(\int_t^u \part_k\gamma^i_0(X_s)(\cald_t X^k_s)ds+
\int_t^u \part_k\gamma^i(X_s)(\cald_t X^k_s)\cdot dW_s\Bigr)
+\gamma^i(X_t)
\ee
it can be shown that $\cald_t X_u$ follows the next SDE:
\bea
d(\cald_{t}X_u^i)&=&\sum_{k=1}^d\Bigl(\part_k\gamma^i_0(X_u)(\cald_t X_u^k)du+\part_k\gamma^i(X_u)(\cald_{t}X_u^k)\cdot dW_u
\Bigr) \\
(\cald_t X^i_t)&=&\gamma^i(X_t)~.
\eea
Thus, comparing to Eqs.~(\ref{YSDE1}) and (\ref{YSDE2}), we can conclude that
\be
(\cald_t X_u)=Y_{t,u} \gamma(X_t)~.
\label{MDY}
\ee
As a result, combining the SDE for $Y_{t,u}$ and 
the Markovian property of $X$, one can confirm that the conditional expectation
\be
E\left[\left.\cald_t
\left(e^{-\int_t^uc(X_s)ds}G(X_u)\right)\right|\calf_t\right]
\ee
is actually given by a some function of $X_t$.
Therefore, in principle, both of the backward components can be expressed in terms of $X_t$ in 
each approximation order. 

In fact, this is an easy task
when the underlying process has a known distribution.
In the next section, we present two such models, and demonstrate how 
our approximation scheme works. We will also compare our approximate solution
to the direct numerical results obtained from,  such as PDE and Monte Carlo simulation.
However, in more generic situations, we do not know the distribution of $X$.
We will explain how to handle the problem in this case using the asymptotic expansion method for the forward components in 
Sec.\ref{asymptotic}.

\section{Simple Examples}
\label{numerical}
\subsection{A forward agreement with bilateral default risk}
As the first example, we consider a toy model for a forward agreement on a stock with 
bilateral default risk of the contracting parties, the investor (party-$1$) and its counterparty (party-$2$).
The terminal payoff of the contract from the view point of the party-$1$ is
\be
\Phi(S_T)=S_T-K
\ee
where $T$ is the maturity of the contract, and $K$ is a constant.
We assume the underlying stock follows a simple geometric Brownian motion:
\be
dS_t=rS_t dt+\sigma S_t dW_t
\ee
where the risk-free interest rate $r$ and the volatility $\sigma$ are assumed to be
positive constants. 
The default intensity of party-$i$ $h_i$ is specified as 
\bea
h_1=\lambda, \hspace{10mm} h_2=\lambda+h
\eea
where $\lambda$ and $h$ are also positive constants.
In this setup, the pre-default value of the contract at time $t$, $V_t$,  
follows~\footnote{See, for example, \cite{Duffie,asymmetric_collateral}.}
\bea
dV_t&=&rV_t dt-h_1\max(-V_t,0)dt+h_2\max(V_t,0)dt+Z_tdW_t\nn \\
&=&(r+\lambda)V_t dt+h\max(V_t,0)dt +Z_t dW_t  \\
V_T&=&\Phi(S_T) ~.
\eea

Now, following the previous arguments, let us introduce the expansion parameter $\ep$, and 
consider the following FBSDE:
\bea
dV_t^{(\ep)}&=&\mu V_t^{(\ep)}dt -\ep g(V_t^{(\ep)})dt+Z_t^{(\ep)}dW_t \\
V_T^{(\ep)}&=&\Phi(S_T) \\
dS_t&=&S_t(rdt+\sigma dW_t)~,
\eea
where we have defined $\mu=r+\lambda$ and $g(v)=-h v\bold{1}_{\{v\geq 0\}}$.
\subsubsection{Zero-th order}
In the zero-th order, we have
\bea
dV_t^{(0)}&=&\mu V_t^{(0)}dt+Z_t^{(0)}dW_t \\
V_T^{(0)}&=&\Phi(S_T)~. 
\eea
Hence we simply obtain
\bea
V_t^{(0)}&=&E\left[\left.e^{-\mu(T-t)}\Phi(S_T)\right|\calf_t\right]\nn \\
&=&e^{-\mu(T-t)}\left(S_t e^{r(T-t)}-K\right)
\eea
and
\be
Z_t^{(0)}=e^{-\lambda(T-t)}\sigma S_t~.
\ee
\subsubsection{First order}
In the first order, we have
\bea
dV_t^{(1)}&=&\mu V_t^{(1)}dt-g(V_t^{(0)})dt+Z_t^{(1)}dW_t \\
V_T^{(1)}&=&0~.
\eea
Thus, we obtain
\bea
V_t^{(1)}&=&E\left[\left.\int_t^T e^{-\mu(T-u)}g(V_u^{(0)})du\right|\calf_t\right] \\
&=&-e^{-\mu(T-t)}h\int_t^T E\left[\left.\max(S_u e^{r(T-u)}-K,0)\right|\calf_t\right] du \\
&=&-e^{-\mu(T-t)}h\int_t^T C(u;t,S_t)du~,
\eea
where
\bea
C(u;t,S_t)&=&S_te^{r(T-t)}N\bigl(d_1(u;t,S_t)\bigr)-KN\bigl(d_2(u;t,S_t)\bigr) \\
d_{1(2)}(u;t,S_t)&=&\frac{1}{\sigma\sqrt{u-t}}\left(\ln\Bigl(\frac{S_t e^{r(T-t)}}{K}\Bigr)\pm
\frac{1}{2}\sigma^2 (u-t)\right)~,
\eea
and $N$ denotes the cumulative distribution function for the standard normal distribution.
We can also derive
\be
Z_t^{(1)}=-e^{-\lambda(T-t)}h\sigma S_t\int_t^T N(d_1(u;t,S_t))du~.
\ee
\subsubsection{Second order}
Finally, let us consider the second order value adjustment.
In this case, the relevant dynamics is given by
\bea
dV_t^{(2)}&=&\mu V_t^{(2)}dt-\frac{\part}{\part v}g(V_t^{(0)})V_t^{(1)}dt+Z_t^{(2)}dW_t\\
V_T^{(2)}&=&0 ~.
\eea
As a result, we have
\bea
V_t^{(2)}&=&E\left[\left.\int_t^T e^{-\mu(u-t)}\left(\frac{\part}{\part v}g(V_u^{(0)})V_u^{(1)}\right)du
\right|\calf_t\right]\\
&=&e^{-\mu(T-t)}h^2 \int_t^T \int_u^T E\left[\left.
\bold{1}_{\{S_ue^{r(T-u)}-K\geq 0\}}C(s;u,S_u)\right|\calf_t\right]ds du
\eea
which can be evaluated as
\bea
V_t^{(2)}=e^{-\mu(T-t)}h^2 \int_t^T \int_u^T \int_{-d_2(u;t,S_t)}^\infty
\phi(z)C\bigl(s;u,S_u(z,S_t)\bigr)dzdsdu~,
\eea
where we have defined
\be
S_u(z,S_t)=S_t e^{\left(r-\frac{1}{2}\sigma^2\right)(u-t)+\sigma\sqrt{u-t}z}
\ee
and 
\be
\phi(z)=\frac{1}{\sqrt{2\pi}}e^{-\frac{1}{2}z^2}~.
\ee
\subsubsection{Numerical comparison to PDE}
For this simple model, we can directly evaluate the contract value $V_t$ by
numerically solving the PDE:
\bea
\frac{\part}{\part v}V(t,S)+\left(
rS\frac{\part}{\part s}V(t,S)+\frac{1}{2}\sigma^2 S^2 \frac{\part^2}{\part s^2}V(t,S)\right)
-\Bigl[\mu+h \bold{1}_{\{V(t,S)\geq 0\}}\Bigr]V(t,S)=0  \nn
\eea
with the boundary conditions
\bea
V(T,S)&=&S-K \nn \\
V(t,M)&=&e^{-(\mu+h)(T-t)}\bigl(Me^{r(T-t)}-K\bigr), \quad M\gg K \nn\\
V(t,m)&=&e^{-\mu(T-t)}\bigl(me^{r(T-t)}-K\bigr), \qquad m\ll K ~.\nn
\eea
\begin{figure}[htbt]
	\center{\includegraphics[width=135mm]{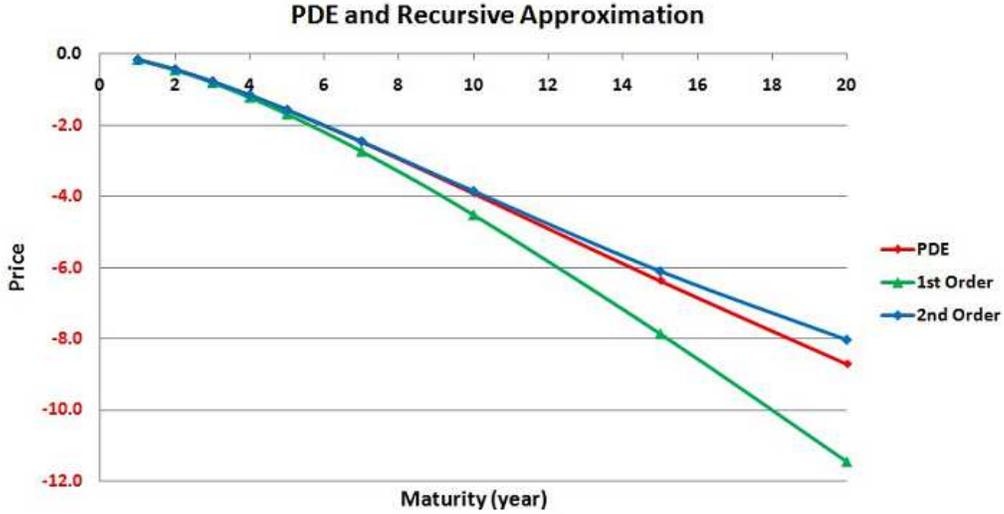}}
	\vspace{-3mm}
	\caption{Numerical Comparison to PDE}
	\label{CVA}
\end{figure} 

In Fig.~\ref{CVA}, we have plot the numerical results of the forward contract with bilateral default risk
with various maturities with the direct solution from the PDE. We have used
\bea
&&r=0.02,\quad \lambda=0.01, \quad h=0.03,\\
&&\sigma=0.2,\quad  S_0=100~,
\eea
where the strike $K$ is chosen to make $V^{(0)}_0=0$ for each maturity.
We have plot $V^{(1)}$ for the first order, and $V^{(1)}+V^{(2)}$ for the 
second order. Note that we have put $\ep=1$ to compare the original model.
One can observe how the higher order correction improves the accuracy of approximation.
In this example, the counterparty is significantly riskier than the investor,
and the underlying contract is quite volatile~\footnote{Of course, people rarely make 
such a risky contract to the counterparty in the real market.}. Even in this situation, the 
simple approximation to the second order works quite well up to the very long maturity.

\subsection{A self-financing portfolio with differential interest rates}
In this subsection, we consider the valuation of self-financing portfolio
under the situation where there exists a difference between the 
lending and borrowing interest rates. Here, we consider the problem 
under the physical measure.

The dynamics of the self-financing portfolio is governed by~\cite{ElKaroui}
\bea
dV_t&=&rV_tdt-\left\{(R-r)\max\left(\frac{Z_t}{\sigma}-V_t,0\right)-\theta Z_t\right\}dt
+Z_t dW_t \\
V_T&=&\Phi(S_T) \\
dS_t&=&S_t\Bigl(\mu dt+\sigma dW_t\Bigr)
\eea
where $r$ and $R$ are the lending and the borrowing rate, respectively.
$\theta=(\mu-r)/\sigma$ denotes the risk premium.
For simplicity, we assume all of the $r$, $R$, $\mu$ and $\sigma$ are positive constants.
Here, $Z_t/\sigma$ represents the amount invested in the risky asset, i.e. stock $S_t$.
Let us choose the terminal wealth function as
\bea
\Phi(S_T)=\max(S_T-K_1,0)-2\max(S_T-K_2,0)~.
\eea
This spread introduces both of the lending and borrowing activities, which makes the problem more interesting.
The setup explained here is in fact exactly the same as that of adopted by
Gobet et al. (2005)~\cite{Gobet}. They have carried our detailed numerical studies for 
the above problem and evaluate $V_0$ by regression-based Monte Carlo simulation.
In the following, we will apply our perturbative approximation scheme to the same problem 
and test its accuracy.

As usual, let us introduce the expansion parameter as
\bea
dV_t^{(\ep)}&=&rV_t^{(\ep)}dt-\ep g(V_t^{(\ep)},Z_t^{(\ep)})dt+Z_t^{(\ep)}dW_t \\
V_T^{(\ep)}&=&\Phi(S_T)~,
\eea
where we have defined the non-linear perturbation function as
\be
g(v,z)=(R-r)\max\left(\frac{z}{\sigma}-v,0\right)-\theta z~.
\ee
Now, we are going to expand $V_t^{(\ep)}$ in terms of $\ep$.

\subsubsection{Zero-th order}
In the zero-th order, the BSDE reduces to 
\bea
dV_t^{(0)}&=&rV_t^{(0)}dt+Z_t^{(0)}dW_t \\
V_T^{(0)}&=&\Phi(S_T)~,
\eea
which allows us to obtain
\bea
V_t^{(0)}&=&E\left[\left. e^{-r(T-t)}\Bigl\{(S_T-K_1)^+-2(S_T-K_2)^+\Bigr\}\right|\calf_t\right] \\
&=&e^{-r(T-t)}\left(C(K_1,S_t)-2C(K_2,S_t)\right)~,
\eea
where we have defined
\bea
C(K_i,S_t)&=&S_te^{\mu(T-t)}N\bigl(d_1(K_i,S_t)\bigr)-K_i N\bigl(d_2(K_i,S_t)\bigr) \\
d_{1(2)}(K_i,S_t)&=&\frac{1}{\sigma\sqrt{T-t}}\left\{
\ln\left(\frac{S_te^{\mu (T-t)}}{K_i}\right)\pm \frac{1}{2}\sigma^2 (T-t)\right\}~
\eea
for $i\in\{1,2\}$.
The volatility term is given by
\be
Z_t^{(0)}=e^{(\mu-r)(T-t)}\sigma S_t\Bigl(N\bigl(d_1(K_1,S_t)\bigr)-2N\bigl(d_2(K_2,S_t)\bigr)\Bigr)~.
\ee
\subsubsection{First order}
Now, in the first order, we have
\bea
dV_t^{(1)}&=&rV_t^{(1)}dt-g(V_t^{(0)},Z_t^{(0)})dt+Z_t^{(1)}dW_t \\
V_T^{(1)}&=&0~.
\eea 
As before, we can easily integrate it to obtain
\bea
V_t^{(1)}=E\left[\left.\int_t^T e^{-r(u-t)}g(V_u^{(0)},Z_u^{(0)})du\right|\calf_t\right]~.
\eea
Now, using the zero-th order results, one can show 
\bea
g(V_t^{(0)},Z_t^{(0)})&=&e^{-r(T-t)}(R-r)\Bigl\{K_1N(d_2(K_1,S_t))-2K_2N(d_2(K_2,S_t))\Bigr\}^+ \nn \\
&-&e^{-r(T-t)}(\mu-r)S_te^{\mu(T-t)}\Bigl\{N(d_1(K_1,S_t))-2N(d_1(K_2,S_t))\Bigr\}~,
\eea
which leads to
\bea
V_t^{(1)}&=&e^{-r(T-t)}\int_t^T du \left\{
E\left[(R-r)\Bigl(K_1N(d_2(K_1,S_u))-2K_2N(d_2(K_2,S_u))\Bigr)^+\right.\right.\nn\\
&&\hspace{20mm}-\left.\left.\left.(\mu-r) S_u e^{\mu(T-u)}\Bigl(N(d_1(K_1,S_u))-2N(d_1(K_2,S_u))\Bigr)\right|\calf_t\right]\right\}~.\nn \\
\eea
By setting 
\be
S_u(S_t,z)=S_te^{\mu(u-t)}\exp\left(-\frac{1}{2}\sigma^2 (u-t)+\sigma\sqrt{u-t}z\right)
\ee
we can write the first order correction as
\bea
&&V_t^{(1)}=e^{-r(T-t)}\int_t^T du\int_{\mathbb{R}}dz \phi(z) \nn \\
&&\hspace{10mm}\left\{(R-r)\Bigl[K_1N(d_2(K_1,S_u(S_t,z)))-2K_2N(d_2(K_2,S_u(S_t,z)))\Bigr]^+ \right.\nn \\
&&\hspace{10mm}-\left.(\mu-r)S_u(S_t,z)e^{\mu(T-u)}\Bigl[
N(d_1(K_1,S_u(S_t,z)))-2N(d_1(K_2,S_u(S_t,z)))\Bigr]\right\}~.\nn \\
\eea
The volatility term can also be derived easily as
\bea
Z_t^{(1)}&=&\sigma S_t \frac{\part}{\part S_t}V_t^{(1)}(S_t)\nn \\
&=&e^{-r(T-t)}\int_t^T du\int_{\mathbb{R}}dz\phi(z)\Bigl[ \nn \\
&&\chi(u,z)\frac{1}{\sqrt{T-u}}\Bigl(K_1\phi(d_2(K_1;S_u(S_t,z)))-2K_2\phi(d_2(K_2,S_u(S_t,z)))\Bigr) \nn \\
&&-(\mu-r)S_u(S_t,z)e^{\mu(T-u)}\Bigl\{
\sigma\Bigl(N(d_1(K_1,S_u(S_t,z)))-2N(d_1(K_2,S_u(S_t,z)))\Bigr)\nn \\
&&\quad +\frac{1}{\sqrt{T-u}}\Bigl(\phi(d_1(K_1,S_u(S_t,z)))-2\phi(d_1(K_2,S_u(S_t,z)))\Bigr) \Bigr\} \Bigr]~,
\eea
where we have defined
\bea
\chi(u,z)&=&1~\quad{\mbox{if}}~\quad K_1N(d_2(K_1,S_u(S_t,z)))-2K_2N(d_2(K_2,S_u(S_t,z)))\geq 0 \nn \\
&=&0~\quad {\mbox{otherwise}}~.
\eea

\subsubsection{Second order}
Finally, in the second order, the relevant FBSDE is given by
\bea
dV_t^{(2)}&=&rV_t^{(2)}dt-\left(\frac{\part}{\part v}g(V_t^{(0)},Z_t^{(0)})V_t^{(1)}
+\frac{\part}{\part z}g(V_t^{(0)},Z_t^{(0)})Z_t^{(1)}\right)dt+Z_t^{(2)}dW_t \nn \\
V_T^{(2)}&=&0~. 
\eea
Using the fact that
\bea
\frac{\part}{\part v}g(v,z)&=&-(R-r)\bold{1}_{\{\frac{z}{\sigma}-v\geq 0\}} \\ 
\frac{\part}{\part z}g(v,z)&=&\frac{R-r}{\sigma}\bold{1}_{\{\frac{z}{\sigma}-v\geq 0\}}-\theta
\eea
we obtain 
\bea
&&V_t^{(2)}=e^{-r(T-t)}\int_t^T du\int_u^T ds \int_{\mathbb{R}}dz_1\int_{\mathbb{R}}dz_2 \phi(z_1)\phi(z_2)\Bigl[ \nn \\
&&\quad\chi(u,z_1)\Bigl\{-(R-r)^2\Bigl(K_1N\bigl(d_2(K_1,S_s(u,\vec{z}))\bigr)-2K_2N\bigl(d_2(K_2,S_s(u,\vec{z}))\bigr)\Bigr)^+ \nn \\
&&+(R-r)(\mu-r)S_s(u,\vec{z})e^{\mu(T-s)}\Bigl(
N(d_1(K_1;S_s(u,\vec{z})))-2N(d_1(K_2,S_s(u,\vec{z})))\Bigr) \Bigr\} \nn \\
&&+\left(\frac{R-r}{\sigma}\chi(u,z_1)-\theta\right)\Bigl\{ \nn \\
&&(R-r)\chi(s,u,\vec{z})\frac{1}{\sqrt{T-s}}\Bigl(K_1\phi(d_2(K_1,S_s(u,\vec{z})))-2K_2\phi(d_2(K_2,S_s(u,\vec{z})))\Bigr) \nn \\
&&-(\mu-r)S_s(u,\vec{z})e^{\mu(T-s)}\Bigl(
\sigma [N(d_1(K_1,S_s(u,\vec{z})))-2N(d_2(K_2,S_s(u,\vec{z})))]\nn\\
&&\hspace{20mm} \left.\left.+\frac{1}{\sqrt{T-s}}[\phi(d_1(K_1,S_s(u,\vec{z})))-2\phi(d_1(K_2,S_s(u,\vec{z}) ))]
\right)\Bigr\}\right]~.
\eea
Here, we have defined
\bea
S_s(u,\vec{z})&=&S_s(S_u(S_t,z_1),z_2) 
\eea
and  also
\bea
\chi(s,u,\vec{z})&=&1 \quad \mbox{if} \quad K_1N(d_2(K_1,S_s(u,\vec{z})))-2K_2N(d_2(K_2,S_s(u,\vec{z})))\geq 0 \nn \\
&=& 0 \quad \mbox{otherwise} ~.
\eea
If one needs, it is also straightforward to derive the volatility component.
\subsubsection{Numerical comparison to the result of Gobet et al.}
Gobet et al. (2005)~\cite{Gobet} have carried out the detailed numerical study
for the above problem using the regression-based Monte Carlo simulation.
They have used
\bea
&&\mu=0.05, \quad \sigma=0.2,\quad r=0.01, \quad R=0.06 \nn \\
&&T=0.25, \quad S_0=100, \quad K_1=95,\quad K_2=105~.
\eea
After trying various sets of basis functions, they have obtained the
price as $V_0=2.95$ with standard deviation $0.01$.

Now, let us provide the results from our perturbative expansion.
We have obtained
\bea
&&V_0^{(0)}=2.7863 \nn \\
&&V_0^{(1)}=0.1814 \nn \\
&&V_0^{(2)}=-0.0149 \nn
\eea
using the same model inputs.
Thus, up to the first order, we have $V_0^{(0)}+V_0^{(1)}=2.968$, which is already fairly close, 
and once we include the second order correction, we have $\sum_{i=0}^2V_0^{(i)}=2.953$, which is 
perfectly consistent with their result of Monte Carlo simulation.
Note that, we have derived analytic formulas with explicit expressions both for the contract value 
and its volatility.

\section{Application of Asymptotic Expansion to Generic Markovian Forward Processes}
\label{asymptotic}
In this section, we consider the situation where the forward components $\{X_t\}$
consist of the generic Markovian processes. In this case, we cannot 
express $V_t^{(i)}$ and $Z_t^{(i)}$ in terms of $\{X_t\}$ exactly,
which prohibits us from obtaining the higher order corrections in a simple fashion 
as we have done in the previous section.

However, notice the fact that what we have to do in each order of expansion is equivalent 
to the pricing of generic European contingent claims and hence we can borrow known techniques
adopted there. In the following, we will explain the use of asymptotic expansion method, now for the 
forward components. Although it is impossible to obtain the exact result, we can still obtain 
analytic expression for $(V_t^{(i)},Z_t^{(i)})$ up to a certain order of the volatilities 
of $\{X_t\}$. For the details of asymptotic expansion for volatility, 
please consult with the works~\cite{KT,STT,T,asymptotic3,TY},
for example.

Let us introduce a new expansion parameter $\delta$, which is now for 
the asymptotic expansion for the forward components.
We express the relevant SDE of generic Markovian process $X^{(\del)}\in \mathbb{R}^d$ as
\bea
dX_u^{(\del)}=\gamma_0(X_u^{(\del)},\del)du+\gamma_a(X_u^{(\del)},\del)dW_u^a~.
\eea
Here, we have used Einstein notation which assumes the summation of all the 
paired indexes. For example, in the above equation, the second term means
\be
\gamma_a(X_u^{(\del)},\del)dW_u^a=\sum_{a=1}^r \gamma_a(X_u^{(\del)},\del)dW_u^a~.
\ee
We assume 
\be
\gamma_a(x,0)=0
\label{delassumption}
\ee
for $a=\{1,\cdots,r\}$. Intuitively speaking,  it suggests that $\del$ counts the order of volatility.

Suppose that, in the $(i-1)$-th order of $\ep$, 
we succeeded to express $V_t^{(i-1)}$ and $Z_t^{(i-1)}$ in 
terms of $X_t^{(\del)}$. Then, in the next order, we can express the backward components as
\bea
V_t^{(i)}&=&E\left[\left.\int_t^T e^{-\int_t^u c(X_s^{(\del)})ds}G(X_u^{(\del)},\del)du\right|\calf_t\right] \\
Z_t^{(i)}&=&E\left[\left.\cald_t
\left(\int_t^T e^{-\int_t^uc(X_s^{(\del)})ds}G(X_u^{(\del)},\del) du \right)\right|\calf_t\right]
\eea
with some function $G$.
If there is no need to obtain $(V_t^{(i+1)},Z_t^{(i+1)})$, we can just run Monte Carlo simulation for $X^{(\del)}$ 
to evaluate these quantities in a standard way.
However, if we want to obtain higher order corrections, we need somehow to express the $(V_t^{(i)},Z_t^{(i)})$
in terms of $X_t^{(\del)}$. 

What we are going to propose here is to expand the backward components around $\del=0$:
\bea
V_t^{(i)}&=&V_t^{(i,0)}+\del V_t^{(i,1)}+\del^2 V_t^{(i,2)}+o(\del^2) \\
Z_t^{(i)}&=&Z_t^{(i,0)}+\del Z_t^{(i,1)}+\del^2 Z_t^{(i,2)}+o(\del^2) 
\eea
and express each $V_t^{(i,j)}$ and $Z_t^{(i,j)}$ in terms of $X_t^{(\del)}$ up to 
a certain order "$j$" of $\del$. Although we can proceed to arbitrarily higher order of $\del$,
we will present explicit expressions up to the second order  in this paper.
For the interested readers, the work~\cite{asymptotic3} provides the systematic methods to 
obtain higher order corrections.
\\

Thank to the well-known chain rule for Malliavin derivative, 
what we have to do is only expanding the two fundamental quantities,
$X_u$ and $\cald_t X_u$ for $u\in[t,T]$, in terms of $\delta$.
Firstly, let us introduce a simpler notation,
\bea
dX_u^{(\del)}&=&\gamma_0(X_u^{(\del)},\del)du+\gamma_a(X_u^{(\del)},\del)dW_u^a \nn \\
&:=&\gamma_\alpha(X_u^{(\del)},\del)dw_u^\alpha~,
\eea
where $\al$ runs through $0$ to $r$ with the convention $w_u^{0}=u$ and $w_u^a=W_u^a$ for $a\in\{1,\cdots,r\}$.
We set the time $t$-value of $X^{(\del)}$ as $x$. Thus our goal is 
to express $V^{(i,j)}$ and $Z^{(i,j)}$ as functions of $x$.
We first introduce a $d\times d$ matrix process $Y^{(\del)}$ defined as
\bea
d(Y_{t,u}^{(\del)})^i_j&=&\part_k\gamma^i_\al(X_u^{(\del)},\del)(Y_{t,u}^{(\del)})^k_jdw_u^\al \\
(Y_{t,t}^{(\del)})^i_j&=&\del^i_j~,
\eea
where $\part_k$ denotes the differential with respect to the $k$-th component of $X$.
Since we have
\bea
(X_u^{(\del)})^i=x^i+\int_t^u \gamma^i_0(X_s^{(\del)},\del)ds+\int_t^u \gamma^i_a(X_s^{(\del)},\del)dW_s^a
\eea
applying a Malliavin derivative $\cald_{t,\beta}$ with $\beta\in\{1,\cdots,r\}$ gives 
\bea
\cald_{t,\beta}(X_u^{(\del)})^i=
\int_t^u \part_k\gamma_0^i(X_s^{(\del)},\del)\cald_{t,\beta}(X_s^{(\del)})^kds
+\int_t^u\part_k\gamma^i_a(X_s^{(\del)},\del)\cald_{t,\beta}(X_s^{(\del)})^kdW_s^a+
\gamma^i_\beta(x,\del)~. \nn
\eea
Thus one can show that
\bea
d\bigl(\cald_{t,\beta}(X_u^{(\del)})^i\bigr)&=&\part_k\gamma^i_\al(X_u^{(\del)},\del)
\cald_{t,\beta}(X_u^{(\del)})^kdw_u^\al \\
\cald_{t,\beta}(X_t^{(\del)})^i&=&\gamma^i_{\beta}(x,\del)~.
\eea
Therefore, for $a\in\{1,\cdots, r\}$, we conclude that
\be
\cald_{t,a} (X_u^{(\del)})^i=(Y^{(\del)}_{t,u})^i_j\gamma^j_a(x,\del)~,
\label{YandM}
\ee
which implies that the asymptotic expansion of $\cald_t X_u^{(\del)}$
can be obtained from that of $Y^{(\del)}$.
Therefore, in the following, we first carry out the asymptotic expansion for $X$ and $Y$.

\subsection{Asymptotic Expansion for $X_u^{(\del)}$ and $Y_u^{(\del)}$}
We are now going to expand for $u\in[t,T]$ as
\be
X_u^{(\del)}=X_u^{(0)}+\delta D_{t,u}+\frac{1}{2}\del^2 E_{t,u}+o(\del^2)~,
\ee
and
\be
Y_{t,u}^{(\del)}=Y_{t,u}+\del H_{t,u}+o(\delta)~,
\ee
where
\bea
D_{t,u}=\left.\frac{\part X_u^{(\del)}}{\part \del}\right|_{\del=0},\qquad 
E_{t,u}=\left.\frac{\part^2 X_u^{(\del)}}{\part \del^2}\right|_{\del=0}~,
\eea
and 
\be
Y_{t,u}=Y_{t,u}^{(0)},\qquad 
H_{t,u}=\left.\frac{\part Y_{t,u}^{(\del)}}{\part \del}\right|_{\del=0}~.
\ee

\subsubsection{Zero-th order}
Since $\gamma_a(\cdot,0)=0$ for $a\in\{1,\cdots,r\}$, we have
\bea
dX_u^{(0)}&=&\gamma_0(X_u^{(0)},0)du \\
d(Y_{t,u})^i_j&=&\part_k\gamma^i_0(X_u^{(0)},0)(Y_{t,u})^k_j du 
\eea
with the initial conditions $X_t^{(0)}=x$ and 
$(Y_{t,t})^i_j=\del^i_j$,
which allows us to express $X_u^{(0)}$ and $Y_{t,u}$ as deterministic functions of $x$.
It is also convenient for later calculations to notice that $Y^{-1}$ is the solution of
\bea
d(Y^{-1}_{t,u})^i_j&=&-(Y_{t,u}^{-1})^i_k\part_j\gamma_0^k(X_u^{(0)},0)du
\label{Yinv}
\eea
with $(Y^{-1}_{t,t})^i_j=\del^i_j$.

\subsubsection{First order}
By applying $\part_\del$, we can easily obtain
\bea
d(\part_\del X_u^{(\del)})^i&=&(\part \gamma_\al(X_u^{(\del)},\del))^i_j\part_\del
(X_u^{(\del)})^jdw_u^\al+\part_\del \gamma^i_\al(X_u^{(\del)},\del)dw_u^\al \\
d(\part_\del Y^{(\del)}_{t,u})^i_j&=&\Bigl\{
\part_k \gamma^i_\al(X_u^{(\del)},\del)(\part_\del Y^{(\del)}_{t,u})^k_j
+\part_{kl}\gamma^i_{\al}(X_u^{(\del)},\del)(\part_\del X_u^{(\del)})^l
(Y_{t,u}^{(\del)})^k_j \nn \\
&&+\part_k\part_\del\gamma^i_\al(X_u^{(\del)},\del)(Y_{t,u}^{(\del)})^k_j\Bigr\}dw_u^\al~. \nn\\
\eea
Putting $\del=0$, they leads to
\bea
dD^{i}_{t,u}&=&\part_j\gamma^i_0(X_u^{(0)},0)D^j_{t,u}du+\part_\del\gamma^i_{\al}(X_u^{(0)},0)dw_u^\al \\
d(H_{t,u})^i_j&=&(\part_k\gamma^i_0(X_u^{(0)},0))(H_{t,u})^k_j du
+\part_{kl}\gamma^i_0(X_u^{(0)},0))D^l_{t,u}(Y_{t,u})^k_j du \nn \\
&&+(\part_k\part_\del\gamma^i_{\al}(X_u^{(0)},0))(Y_{t,u})^k_j dw_u^\al~.
\eea
Now, by using Eq.(\ref{Yinv}), one can show that
\bea
&&D^i_{t,u}=(Y_{t,u})^i_j\int_t^u(Y_{t,s}^{-1})^j_k\part_\del\gamma_\al^k(s)dw_s^\al \\
&&(H_{t,u})^i_j=(Y_{t,u})^i_k\int_t^u(Y^{-1}_{t,s})^k_l
\Bigl\{(\part_{mn}\gamma^l_0(s))D^n_{t,s}(Y_{t,s})^m_jds+
(\part_m\part_\del \gamma^l_\al(s))(Y_{t,s})^m_jdw_s^\al\Bigr\}\nn \\
\eea
where we have defined the shorthand notation that
\be
\gamma^i_j(s):=\gamma^i_j(X_s^{(0)},0),
\ee
which will be used in the following calculations, too.
\subsubsection{Second order}
Applying $\part_\del^2$ to the SDE of $X^{(\del)}$ gives us
\bea
d(\part_\del^2 X_u^{(\del)})^i&=&\Bigl\{
(\part \gamma_\al(X_u^{(\del)},\del))^i_j\part_\del^2 (X_u^{(\del)})^j
+\part_{jk}\gamma^i_\al(X_u^{(\del)},\del)\part_\del(X_u^{(\del)})^j\part_\del (X_u^{(\del)})^k\nn \\
&&+2\part_j\part_\del \gamma^i_\al(X_u^{(\del)},\del)\part_\del (X_u^{(\del)})^j+
\part_\del^2 \gamma^i_\al(X_u^{(\del)},\del)\Bigr\}dw_u^\al~.
\eea
Thus, putting $\del=0$, we obtain
\bea
dE^{i}_{t,u}&=&(\part \gamma_0(X_u^{(0)},0))^i_jE^j_{t,u}du+\part_{jk}\gamma^i_0(X_u^{(0)},0)D^j_{t,u}D^k_{t,u}du\nn \\
&&+2\part_j\part_\del\gamma^i_\al(X_u^{(0)},0)D^j_{t,u}dw_u^\al+\part_\del^2\gamma^i_\al(X_u^{(0)},0)dw_u^\al~.
\eea
Now we can integrate it as
\bea
E^i_{t,u}=(Y_{t,u})^i_j\int_t^u(Y^{-1}_{t,s})^j_k\Bigl\{
\part_{lm}\gamma^k_0(s)D^l_{t,s}D^m_{t,s}ds+\Bigl(2\part_l\part_\del\gamma^k_\al(s)D^l_{t,s}
+\part_\del^2\gamma^k_\al(s)\Bigr)dw_s^\al\Bigr\}.~\nn\\
\eea
We do not need the second order terms for $Y^{(\del)}$.

\subsection{Asymptotic Expansion for Malliavin Derivative:~$\cald_t X_u^{(\del)}$}
For convenience, let us define
\be
(\calx_a^i)^{(\del)}_{t,u}=(\cald_t X_u^{(\del)})^i_a
\ee
and its expansion as
\be
(\calx_a^i)^{(\del)}_{t,u}=\del(\calx_a^i)^{(1)}_{t,u}+\frac{1}{2}\del^2 (\calx_a^i)^{(2)}_{t,u}+o(\del^2)~,
\ee
where
\bea
(\calx_a^i)_{t,u}^{(1)}=\left.\frac{\part}{\part \del} (\calx_a^i)^{(\del)}_{t,u}\right|_{\del=0},\qquad
(\calx_a^i)_{t,u}^{(2)}=\left.\frac{\part^2}{\part \del^2}(\calx_a^i)^{(\del)}_{t,u}\right|_{\del=0}~.
\eea
Note that, the zero-th order term $(\calx_a^i)^{(0)}$ vanishes, due to the assumption (\ref{delassumption}).

From (\ref{YandM}), we can easily show that
\bea
(\calx_a^i)^{(1)}_{t,u}&=&(Y_{t,u})^i_j(\part_\del \gamma^j_a(x,0)) \\
(\calx_a^i)^{(2)}_{t,u}&=&(Y_{t,u})^i_j(\part_\del^2 \gamma^j_a(x,0))+
2(H_{t,u})^i_j(\part_\del \gamma^j_a(x,0))~.
\eea

\subsection{Asymptotic Expansion for  $V^{(i,\del)}$}
Now, we try to express
\bea
V_t^{(i,\del)}=\int_t^T E\left[\left.e^{-\int_t^u c(X_s^{(\del)})ds}G(X_u^{(\del)},\del)\right|\calf_t\right]du
\eea
as a function of $x=X_t^{(\del)}$ using the previous results.
For that purpose, we first need to carry out asymptotic expansion for 
\be
R_{t,u}^{(\del)}:=e^{-\int_t^u c(X_s^{(\del)})ds}G(X_u^{(\del)},\del)~
\ee
to obtain
\bea
R^{(\del)}_{t,u}=R^{(0)}_{t,u}+\del R^{(1)}_{t,u}+\frac{1}{2}\del^2 R^{(2)}_{t,u}+o(\del^2)~,
\eea
where 
\bea
R^{(1)}_{t,u}=\left.\frac{\part}{\part \del}R^{(\del)}_{t,u}\right|_{\del=0}~,\qquad 
R^{(2)}_{t,u}=\left.\frac{\part^2}{\part \del^2}R^{(\del)}_{t,u}\right|_{\del=0}~.
\eea
Then, we can take the conditional expectation straightforwardly.

\subsubsection{Zero-th order}
We have
\be
R^{(0)}_{t,u}=e^{-\int_t^u c(X_s^{(0)})ds}G(X_u^{(0)},0)
\ee
which is a deterministic function of $x$.

\subsubsection{First order}
It is easy to obtain
\bea
\part_\del R^{(\del)}_{t,u}&=&e^{-\int_t^u c(X_s^{(\del)})ds}
\Bigl\{-G(X_u^{(\del)},\del)\int_t^u \part_i c(X_s^{(\del)})(\part_\del X_s^{(\del)})^ids \nn \\
&&\qquad +(\part_i G(X_u^{(\del)},\del))(\part_\del X_u^{(\del)})^i+\part_\del G(X_u^{(\del)},\del)\Bigr\}~,
\eea
which leads to 
\bea
R^{(1)}_{t,u}=e^{-\int_t^u c(X_s^{(0)})ds}\Bigl\{
(\part_i G(X_u^{(0)},0))D_{t,u}^i+\part_\del G(X_u^{(0)},0)-
G(X_u^{(0)},0)\int_t^u\part_i c(X_s^{(0)})D^i_{t,s}ds\Bigr\}.\nn \\
\eea
\subsubsection{Second order}
In the same way, we can show that
\bea
R^{(2)}_{t,u}&=&e^{-\int_t^u c(X_s^{(0)})ds}\Bigl\{(\part_{ij}G(X_u^{(0)},0))D^i_{t,u}D^j_{t,u}+2(\part_i\part_\del G(X_u^{(0)},0))D^i_{t,u} \nn \\
&&\qquad +(\part_i G(X_u^{(0)},0))E^i_{t,u}+\part_\del^2 G(X_u^{(0)},0) \nn \\
&&\qquad +\left(\int_t^u \part_ic(X_s^{(0)})D^i_{t,s}ds\right)^2G(X_u^{(0)},0) \nn \\
&&\qquad -2\left(\int_t^u \part_i c(X_s^{(0)})D^i_{t,s}ds\right)\Bigl(\part_j G(X_u^{(0)},0)D^j_{t,u}
+\part_\del G(X_u^{(0)},0) \Bigr)\nn \\
&&\qquad-\left(\int_t^u\bigl[\part_{ij}c(X_s^{(0)})D^i_{t,s}D^j_{t,s}+\part_i c(X_s^{(0)})E^i_{t,s}\Bigr]ds
\right)G(X_u^{(0)},0)\Bigr\}~.
\eea
\subsubsection{Expression for $V_t^{(i,\del)}$}
Evaluation of the conditional expectation can be easily done 
by simply applying Ito-isometry.
Let us first define
\bea
\ol{D}^i_{t,u}&=&(Y_{t,u})^i_j\int_t^u (Y_{t,s}^{-1})^j_k(\part_\del \gamma^k_0(s))ds \\
\hat{D}^i_{t,u}&=&(Y_{t,u})^i_j\int_t^u (Y_{t,s}^{-1})^j_k(\part_\del \gamma^k_a(s))dW^a_s
\eea
and then we have $D^i_{t,u}=\ol{D}^i_{t,u}+\hat{D}^i_{t,u}$.
Since the first one is a deterministic function, we have,
for $\forall u, s\geq t$,
\bea
\ol{D^i_{t,u}D^j_{t,s}}&:=&E\left[\left. D^i_{t,u}D^j_{t,s}\right|\calf_t\right] \nn \\
&=&\ol{D}^i_{t,u}\ol{D}^j_{t,s}+\ol{\hat{D}^i_{t,u}\hat{D}^j_{t,s}}~,
\eea
where
\be
\ol{\hat{D}^i_{t,u}\hat{D}^j_{t,s}}=
(Y_{t,u})^i_k(Y_{t,s})^j_l\int_t^{u\wedge s}
(Y^{-1}_{t,v})^k_m(Y^{-1}_{t,v})^l_n
(\part_\del \gamma^m_a(v))(\part_\del \gamma^n_a(v))dv~.
\ee
Then, similarly, we can express
\bea
&&\ol{E}_{t,u}^i:=E\left[\left.E_{t,u}^i\right|\calf_t\right]\nn \\
&&\quad=(Y_{t,u})^i_j\int_t^u (Y_{t,s}^{-1})^j_k
\Bigl\{\part_{lm}\gamma^k_0(s)\ol{D^l_{t,s}D^m_{t,s}}
+2\part_l\part_\del \gamma^k_0(s)\ol{D}^l_{t,s}+\part_\del^2\gamma^k_0(s)\Bigr\}ds~.
\eea
\\

Using these results, we have
\bea
&&\ol{R}^{(0)}_{t,u}:=E\left[\left.R_{t,u}^{(0)}\right|\calf_t\right]=e^{-\int_t^u c(X_s^{(0)})ds}G(X_u^{(0)},0) \\
&&\ol{R}^{(1)}_{t,u}:=E\left[\left.R_{t,u}^{(1)}\right|\calf_t\right]\nn \\
&&~~=e^{-\int_t^u c(X_s^{(0)})ds}\Bigl\{
(\part_i G(X_u^{(0)},0))\ol{D}^i_{t,u}+\part_\del G(X_u^{(0)},0)
-G(X_u^{(0)},0)\int_t^u \part_i c(X_s^{(0)})\ol{D}^i_{t,s}ds\Bigr\}~,\nn \\
\eea
and also
\bea
&&\ol{R}^{(2)}_{t,u}:=E\left[\left.R_{t,u}^{(2)}\right|\calf_t\right]=e^{-\int_t^u c(X_s^{(0)})ds}\Bigl\{
\nn \\
&&\qquad \part_{ij}G(X_u^{(0)},0)\ol{D^i_{t,u}D^j_{t,u}}+
2(\part_i\part_\del G(X_u^{(0)},0))\ol{D}^i_{t,u}+\part_i G(X_u^{(0)},0)\ol{E}^i_{t,u}
+\part_\del^2 G(X_u^{(0)},0)\nn \\
&&\qquad+G(X_u^{(0)},0)\int_t^u\int_t^u\part_i c(X_s^{(0)})\part_jc(X_v^{(0)})\ol{D^i_{t,s}D^j_{t,v}}dsdv\nn \\
&&\qquad -2\part_jG(X_u^{(0)},0)\int_t^u
\part_i c(X_s^{(0)})\ol{D^i_{t,s}D^j_{t,u}}ds \nn \\
&&\qquad -2\part_\del G(X_u^{(0)},0)\int_t^u \part_i c(X_s^{(0)})\ol{D}^i_{t,s}ds \nn \\
&&\qquad -G(X_u^{(0)},0)\int_t^u
\bigl[ \part_{ij}c(X_s^{(0)})\ol{D^i_{t,s}D^j_{t,s}}+\part_ic(X_s^{(0)})\ol{E}^i_{t,s}\bigr]ds\Bigr\}.
\eea
Now, we are able to express $V_t^{(i,\del)}$ as a function of $x$ up to the second order of $\del$ as
desired:
\bea
V_t^{(i,\del)}=\int_t^T \Bigl\{
\ol{R}_{t,u}^{(0)}+\del \ol{R}^{(1)}_{t,u}+
\frac{1}{2}\del^2~\ol{R}^{(2)}_{t,u}\Bigr\}du +o(\del^2)~.
\eea

\subsection{Asymptotic Expansion for  $Z^{(i,\del)}$}
Finally, we are going to express 
\be
Z_t^{(i,\del)}=\int_t^T E\left[\left.
\cald_t\left(e^{-\int_t^u c(X_s^{(\del)})ds}G(X_u^{(\del)},\del)\right)\right|\calf_t\right]du
\ee
as a function of $x=X_t^{(\del)}$.
Let us introduce the two quantities:
\bea
(\eta_a^{(\del)})_{t,u}&=&e^{-\int_t^u c(X_s^{(\del)})ds}\part_i G(X_u^{(\del)},\del)(\cald_t X_u^{(\del)})^i_a \\
(\xi_a^{(\del)})_{t,u}&=&-e^{-\int_t^uc(X_s^{(\del)})ds}G(X_u^{(\del)},\del)
\left(\int_t^u \part_i c(X_s^{(\del)})(\cald_t X_s^{(\del)})^i_a ds\right).
\eea
Then, we have
\be
\cald_t\left(e^{-\int_t^u c(X_s^{(\del)})ds}G(X_u^{(\del)},\del)\right)=(\eta_a^{(\del)})_{t,u}
+(\xi_a^{(\del)})_{t,u}.
\ee
Similarly to the previous section, we try to obtain the expressions as
\bea
(\eta_a^{(\del)})_{t,u}&=&\del (\eta_a^{(1)})_{t,u}+\frac{1}{2}\del^2 (\eta^{(2)}_a)_{t,u}+o(\del^2) \\
(\xi_a^{(\del)})_{t,u}&=&\del (\xi_a^{(1)})_{t,u}+\frac{1}{2}\del^2 (\xi^{(2)}_a)_{t,u}+o(\del^2) 
\eea
where both of the zero-th order terms vanish.

We have
\bea
&&\hspace{-10mm}\part_\del (\eta^{(\del)}_a)_{t,u}=e^{-\int_t^u c(X_s^{(\del)})ds}\Bigl\{
-\left(\int_t^u \part_i c(X_s^{(\del)})(\part_\del X_s^{(\del)})^ids\right)
\part_jG(X_u^{(\del)},\del)(\cald_t X_u^{(\del)})^j_a \nn \\
&&\hspace{-10mm}~+(\part_{ij}G(X_u^{(\del)},\del))(\part_\del X_u^{(\del)})^j(\cald_t X_u^{(\del)})^i_a
+(\part_i\part_\del G(X_u^{(\del)},\del))(\cald_t X_u^{(\del)})_a^i+
(\part_i G(X_u^{(\del)},\del))(\part_\del \cald_t X_u^{(\del)})^i_a \Bigr\}. \nn \\
\eea
Thus, we obtain
\bea
(\eta_a^{(1)})_{t,u}&:=&\left.\frac{\part}{\part \del}(\eta_a^{(\del)})_{t,u}\right|_{\del=0} \nn \\
&=&e^{-\int_t^u c(X_s^{(0)})ds}\part_iG(X_u^{(0)},0)(\calx_a^i)^{(1)}_{t,u}~.
\eea
Similarly we can show that
\bea
(\eta^{(2)}_a)_{t,u}&:=&\left.\frac{\part^2}{\part \del^2}(\eta^{(\del)}_a)_{t,u}\right|_{\del=0}\nn\\
&=&-2\left(\int_t^u \part_ic(X_s^{(0)})D_{t,s}^ids\right)(\eta_a^{(1)})_{t,u}\nn\\
&&\hspace{-18mm}+e^{-\int_t^u c(X_s^{(0)})ds}\Bigl[
(\part_iG(X_u^{(0)},0))(\calx_a^i)^{(2)}_{t,u}+2(\part_{ij}G(X_u^{(0)},0))D_{t,u}^j(\calx_a^i)^{(1)}_{t,u}
+2(\part_i\part_\del G(X_u^{(0)},0))(\calx_a^i)^{(1)}_{t,u}\Bigr]~.\nn\\
\eea

In the same way, for $\xi^{(\del)}$, we have
\bea
&&\part_\del(\xi^\del_a)_{t,u}=e^{-\int_t^u c(X_s^{(\del)})ds}G(X_u^{(\del)},\del)
\left(\int_t^u (\part_ic(X_s^{(\del)}))(\part_\del X_s^{(\del)})^ids\right)
\left(\int_t^u (\part_j c(X_s^{(\del)}))(\cald_t X_s^{(\del)})^j_a ds\right)\nn\\
&&-e^{-\int_t^u c(X_s^{(\del)})ds}\Bigl\{\part_\del G(X_u^{(\del)},\del)+(\part_i G(X_u^{(\del)},\del))(\part_\del X_u^{(\del)})^i\Bigr\}\left(\int_t^u \part_i c(X_s^{(\del)})(\cald_t X_s^{(\del)})_a^i ds\right)\nn\\
&&-e^{-\int_t^u c(X_s^{(\del)})ds}G(X_u^{(\del)},\del)
\int_t^u\left[
(\part_{ij}c(X_s^{(\del)}))(\part_\del X_s^{(\del)})^j(\cald_t X_s^{(\del)})^i_a
+(\part_i c(X_s^{(\del)}))(\part_\del \cald_t X_s^{(\del)})^i_a\right]ds~. \nn\\
\eea
Thus we can show that
\bea
(\xi^{(1)}_a)_{t,u}&:=&\left.\frac{\part}{\part \del}(\xi^\del_a)_{t,u}\right|_{\del=0}\nn\\
&=&-e^{-\int_t^u c(X_s^0)ds}G(X_u^{(0)},0)\int_t^u(\part_i c(X_s^{(0)}))(\calx_a^i)^{(1)}_{t,s}ds
\eea
and similarly
\bea
&&(\xi^{(2)}_a)_{t,u}=\left. \frac{\part^2}{\part \del^2}(\xi^\del_a)_{t,u}\right|_{\del=0}\nn \\
&&=-2\left(\int_t^u (\part_i c(X_s^{(0)}))D_{t,s}^i ds\right)(\xi^{(1)}_a)_{t,u}\nn\\
&&-e^{-\int_t^u c(X_s^{(0)})ds}\Bigl\{
2\left(\part_\del G(X_u^{(0)},0)+(\part_i G(X_u^{(0)},0) )D_{t,u}^i\right)
\int_t^u (\part_j c(X_s^{(0)}) )(\calx^j_a)^{(1)}_{t,s}ds \nn \\
&&\quad+G(X_u^{(0)},0)\int_t^u\Bigl[
(\part_i c(X_s^{(0)}) )(\calx_a^i)^{(2)}_{t,s}+
2(\part_{ij}c(X_s^{(0)}))D_{t,s}^j(\calx_a^i)^{(1)}_{t,s}\Bigr]ds \Bigr\}~.
\eea

For the evaluation of the conditional expectation, let us define, for $\forall u\geq t$, 
\bea
&&\hspace{-18mm}(\ol{H}_{t,u})^i_j:=E\left[\left.(H_{t,u})^i_j\right|\calf_t\right]\nn \\
&&\hspace{-18mm}\quad=(Y_{t,u})^i_k\int_t^u (Y^{-1}_{t,s})^k_l\Bigl\{ (\part_{mn}\gamma^l_0(s))\ol{D}^n_{t,s}+
(\part_m\part_\del \gamma^l_0(s))\Bigr\}(Y_{t,s})^m_jds \\
&&\hspace{-18mm}(\ol{\calx_a^i})^{(1)}_{t,u}:=E\left[\left.(\calx_a^i)^{(1)}_{t,u}\right|\calf_t\right]=(Y_{t,u})^i_j(\part_\del \gamma^j_a(x,0))~\\
&&\hspace{-18mm}(\ol{\calx}_a^i)^{(2)}_{t,u}:=E\left[\left.(\calx_a^i)^{(2)}_{t,u}\right|\calf_t\right]=(Y_{t,u})^i_j(\part_\del^2 \gamma^j_a(x,0))+2(\ol{H}_{t,u})^i_j(\part_\del \gamma^j_a(x,0)).
\eea
Using these expressions, one show that
\bea
(\ol{\eta}^{(1)}_a+\ol{\xi}^{(1)}_a)_{t,u}&:=&E\left[\left. (\eta^{(1)}_a+\xi^{(1)}_a)_{t,u}\right|\calf_t\right]\nn \\
&&\hspace{-25mm}=e^{-\int_t^u c(X_s^{(0)})ds}\Bigl\{
(\part_i G(X_u^{(0)},0) )(\ol{\calx}_a^i)^{(1)}_{t,u}
-G(X_u^{(0)},0)\int_t^u (\part_i c(X_s^{(0)}) )(\ol{\calx}_a^i)^{(1)}_{t,s}ds\Bigr\}, \nn\\
\eea
and in the same way that
\bea
&&\hspace{-5mm}(\ol{\eta}^{(2)}_a
+\ol{\xi}^{(2)}_a)_{t,u}:=E\left[\left. (\eta^{(2)}_a+\xi^{(2)}_a)_{t,u}\right|\calf_t\right]\nn\\
&&=-2\left(\int_t^u (\part_i c(X_s^{(0)}))\ol{D}_{t,s}^ids\right)(\ol{\eta}^{(1)}_a+\ol{\xi}^{(1)}_a)_{t,u} \nn\\
&&+e^{-\int_t^u c(X_s^{(0)})ds}\Bigl\{
(\part_i G(X_u^{(0)},0))(\ol{\calx}_a^i)^{(2)}_{t,u}+2(\part_{ij}G(X_u^{(0)},0))\ol{D}^j_{t,u}(\ol{\calx}_a^i)^{(1)}_{t,u}
+2(\part_i\part_\del G(X_u^{(0)},0))(\ol{\calx}_a^i)^{(1)}_{t,u}\nn\\
&&\quad -2\left[\part_\del G(X_u^{(0)},0)+(\part_i G(X_u^{(0)},0))\ol{D}^i_{t,u}\right]\int_t^u
(\part_j c(X_s^{(0)}))(\ol{\calx}_a^j)^{(1)}_{t,s}ds \nn \\
&&\quad -G(X_u^{(0)},0)\int_t^u\Bigl[
(\part_i c(X_s^{(0)}) )(\ol{\calx}_a^i)^{(2)}_{t,s}+
2(\part_{ij}c(X_s^{(0)}))\ol{D}_{t,s}^j(\ol{\calx}_a^i)^{(1)}_{t,s}\Bigr]ds~\Bigr\}.
\eea
Now that we are able to express $Z_t^{(i,\del)}$ as a function of $x=X_t^{(\del)}$ as
\bea
(Z_a^{(i,\del)})_t=\int_t^T \Bigl\{ \del(\ol{\eta}_a^{(1)}+\ol{\xi}^{(1)}_a)_{t,u}
+\frac{1}{2}\del^2 (\ol{\eta}^{(2)}_a+\ol{\xi}^{(2)}_a)_{t,u}\Bigr\}du+o(\del^2)~.
\eea
\\

This completes the goal of asymptotic expansion for $V^{(i,\del)}$ and $Z^{(i,\del)}$, which 
are now expressed as functions of $x$ as desired.

\section{Perturbation in PDE Framework}
In this section, we will study the perturbation scheme under the PDE (partial differential equation) framework
following the so-called four step scheme~\cite{fourstep}.
We will see that our perturbative method makes the four step scheme tractable for the generic situations,
which only requires standard techniques for the classical parabolic linear PDE.
In the next section, we will explain the equivalent perturbation method 
in the probabilistic framework.

\subsection{PDE Formulation based on Four Step Scheme}
Let us consider the following generic coupled non-linear FBSDE:
\bea
dV_t&=&-f(t,X_t,V_t,Z_t)dt+Z_t\cdot dW_t \nn \\
V_T&=&\Phi(X_T) \nn \\
dX_t&=&\gamma_0(t,X_t,V_t,Z_t)dt+\gamma(t,X_t,V_t,Z_t)\cdot dW_t \nn\\
X_0&=&x~.
\label{fbsde_org}
\eea
Here, we made the dependence on $t$ explicitly  to clearly distinguish it from the stochastic $X$ components.
As before,  we assume that $V$, $Z$, $X$ take value in $\mathbb{R}$, $\mathbb{R}^r$ and $\mathbb{R}^d$ respectively,
and $W$ denotes a $r$ dimensional standard Brownian motion.

Following the arguments of the four step scheme of Ma and Yong~\cite{Ma}, let us postulate that $V_t$ is 
given by the function of $t$ and $X_t$ as
\be
V_t=v(t,X_t)
\ee
almost surely for $\forall t\in[0,T]$. Then, applying It$\hat{o}$'s formula, we obtain
\bea
dV_t&=& \part_t v(t,X_t) dt\nn \\
&&+\Bigl\{\part_i v(t,X_t)\gamma_0^i(t,X_t,v(t,X_t),Z_t)+\frac{1}{2}\part_{ij}v(t,X_t)(\gamma^i\cdot \gamma^j)
(t,X_t,v(t,X_t),Z_t)\Bigr\}dt\nn \\
&&+\part_i v(t,X_t)\gamma^i (t,X_t,v(t,X_t),Z_t)\cdot dW_t~.
\eea
Thus, in order that $v$ is the right choice, it should satisfy
\bea
\label{T_pde}
&&\hspace{-12mm}v(T,x)=\Phi(x) \\
\label{pde_org}
&&\hspace{-12mm}\part_t v(t,x)+\Bigl\{
\part_i v(t,x)\gamma^i_0(t,x,v(t,x),z(t,x))+\frac{1}{2}\part_{ij}v(t,x) (\gamma^i\cdot \gamma^j)(t,x,v(t,x),z(t,x))\Bigr\} \nn \\
&&+f(t,x,v(t,x),z(t,x))=0 \\
\label{z_pde}
&&\hspace{-12mm}z(t,x)=\part_iv(t,x)\gamma^i(t,x,v(t,x),z(t,x))~,
\eea
where the last equation arises to match the volatility term.
\\

In the four step scheme, one first needs to find the solution $z(t,x)$ satisfying the Eq.(\ref{z_pde}). And secondly,
one has to solve the PDE (\ref{pde_org}) to obtain $v(t,x)$, which then allows one to run $X$ as a standalone
Markovian process in the third step. 
And then finally, one will obtain the backward components by setting $V_t=v(t,X_t)$ and $Z_t=z(t,X_t)$. 
The crucial point in the above four step scheme is whether one can finish the step $1$ and $2$ successfully.
Even if one finds the solution for $z$, the second step requires 
to solve the non-linear PDE (\ref{pde_org}), which is very difficult in general. 
In the remainder of this section, let us study how our perturbation method works to achieve this goal.
\\

We consider, as before, the original system Eq~(\ref{fbsde_org}) as a linear decoupled FBSDE with 
perturbations of non-linear generator and feedbacks of the order of $\ep$. We write it as
\bea
&&dV_t^{(\ep)}=c(t,X^{(\ep)}_t)V_t^{(\ep)}dt-\ep g(t,X_t^{(\ep)},V_t^{(\ep)},Z_t^{(\ep)})dt+Z_t^{(\ep)}\cdot dW_t \nn \\
&&V_T^{(\ep)}=\Phi(X_T^{(\ep)})\nn \\
&&dX_t^{(\ep)}=\Bigl(r(t,X_t^{(\ep)})+\ep\mu(t,X_t^{(\ep)},V_t^{(\ep)},Z_t^{(\ep)})\Bigr)dt \nn \\
&&\hspace{25mm}+\Bigl(\sigma(t,X_t^{(\ep)})+\ep \eta(t,X_t^{(\ep)},V_t^{(\ep)},Z_t^{(\ep)})\Bigr)\cdot dW_t \nn \\
&&X_0^{(\ep)}=x~
\label{p_fbsde}
\eea
and the corresponding PDE:
\bea
\label{p_pde}
&&\hspace{-7mm}v^{(\ep)}(T,x)=\Phi(x) \nn \\
&&\hspace{-7mm}\part_t v^{(\ep)}(t,x)+\Bigl\{
\part_i v^{(\ep)}(t,x)\gamma^{i}_0(t,x,v^{(\ep)},z^{(\ep)})+\frac{1}{2}\part_{ij}v^{(\ep)}(t,x) (\gamma^{i}\cdot \gamma^{j})(t,x,v^{(\ep)},z^{(\ep)})\Bigr\}\nn \\
&&\hspace{0mm}+f(t,x,v^{(\ep)},z^{(\ep)})=0 \nn \\
&&\hspace{-7mm}z^{(\ep)}(t,x)=\part_i v^{(\ep)}(t,x)\gamma^{i}(t,x,v^{(\ep)}(t,x),z^{(\ep)}(t,x))~,
\eea
where
\bea
f(t,x,v^{(\ep)},z^{(\ep)})&=&-c(t,x)v^{(\ep)}(t,x)+\ep g(t,x,v^{(\ep)}(t,x),z^{(\ep)}(t,x)) \\
\gamma_0(t,x,v^{(\ep)},z^{(\ep)})&=&r(t,x)+\ep\mu(t,x,v^{(\ep)}(t,x),z^{(\ep)}(t,x))\\
\gamma(t,x,v^{(\ep)},z^{(\ep)})&=&\sigma(t,x)+\ep\eta(t,x,v^{(\ep)}(t,x),z^{(\ep)}(t,x)) ~.
\eea
We suppose that the solution of the above PDE can be expanded perturbatively in such a way that
\bea
v^{(\ep)}(t,x)&=&v^{(0)}(t,x)+\ep v^{(1)}(t,x)+\ep^2 v^{(2)}(t,x)+\cdots \\
z^{(\ep)}(t,x)&=&z^{(0)}(t,x)+\ep z^{(1)}(t,x)+\ep^2 z^{(2)}(t,x)+\cdots ~,
\eea
and then try to solve $v^{(i)},z^{(i)}$ order by order.
If the non-linear terms are small enough, 
we can expect to obtain a good approximation by putting $\ep=1$ in the above expansion 
to a certain order.

\subsection{Zero-th order}
In the zero-th order, the PDE (\ref{p_pde}) reduces to
\bea
&\bigl(\part_t +\call(t,x)\bigr)v^{(0)}(t,x)=0& \nn \\
&v^{(0)}(T,x)=\Phi(x)&
\label{pde_v0}
\eea
and 
\be
z^{(0)}(t,x)=\part_i v^{(0)}(t,x)\sigma^i(t,x)~.
\label{pde_z0}
\ee
Here, we have defined the operator $\call$ as
\be
\call(t,x)=r^i(t,x)\part_i+\frac{1}{2}(\sigma^i\cdot \sigma^j)(t,x)\part_{ij}-c(t,x)~.
\ee
This is a standard parabolic PDE and can be handled in the usual way.
One can easily check that $V_t=v^{(0)}(t,X_t)$ and $Z_t=z^{(0)}(t,X_t)$ solves 
the FBSDE (\ref{p_fbsde}) when $\ep=0$.

\subsection{First order}
By extracting $\ep$-first order terms from the PDE, we obtain
\bea
&\bigl(\part_t+\call(t,x)\bigr)v^{(1)}(t,x)+G^{(1)}(t,x)=0& \nn \\
&v^{(1)}(T,x)=0&
\label{pde_v1}
\eea
and 
\be
z^{(1)}(t,x)=\part_i v^{(1)}(t,x)\sigma^i(t,x)+\part_i v^{(0)}(t,x)\eta^{i(0)}(t,x)~.
\label{pde_z1}
\ee
Here, we have defined
\be
G^{(1)}(t,x)=\part_i v^{(0)}(t,x)\mu^{i(0)}(t,x)+\part_{ij}v^{(0)}(t,x)
(\sigma^i\cdot\eta^{j(0)})(t,x)+g^{(0)}(t,x)
\ee
and the following notations:
\bea
\mu^{(0)}(t,x)&=&\mu(t,x,v^{(0)}(t,x),z^{(0)}(t,x)) \\
\eta^{(0)}(t,x)&=&\eta(t,x,v^{(0)}(t,x),z^{(0)}(t,x))\\
g^{(0)}(t,x)&=&g(t,x,v^{(0)}(t,x),z^{(0)}(t,x))~.
\eea
As a result, we once again obtained a linear parabolic PDE. Hence $(v^{(1)},z^{(1)})$ can also be solved, 
at least numerically, in a standard fashion.

\subsection{Second order}
In the second order, one can show $v^{(2)}$ and $z^{(2)}$ should satisfy
\bea
& \bigl(\part_t +\call(t,x)\bigr)v^{(2)}(t,x)+G^{(2)}(t,x)=0 & \nn \\
&v^{(2)}(T,x)=0& 
\label{pde_v2}
\eea
and 
\bea
z^{(2)}(t,x)&=&\part_i v^{(2)}(t,x)\sigma^i(t,x)+\part_i v^{(1)}(t,x)\eta^{i(0)}(t,x)\nn \\
&&+\part_i v^{(0)}(t,x)\bigl(v^{(1)}(t,x)\part_v+z^{(1)}(t,x)\cdot \nabla_z\bigr)\eta^{i(0)}(t,x)~.
\label{pde_z2}
\eea
Here, $G^{(2)}$ is given by
\bea
G^{(2)}(t,x)&=&
\part_i v^{(1)}(t,x)\mu^{i(0)}(t,x)+\part_i v^{(0)}(t,x)
\bigl(v^{(1)}(t,x)\part_v+z^{(1)}(t,x)\cdot \nabla_z\bigr)\mu^{i(0)}(t,x) \nn \\
&&+\part_{ij}v^{(1)}(t,x)(\sigma^i\cdot \eta^{j(0)})(t,x)+\frac{1}{2}\part_{ij}v^{(0)}(t,x)
(\eta^{i(0)}\cdot \eta^{j(0)})(t,x)\nn \\
&&+\part_{ij}v^{(0)}(t,x)\sigma^{i}(t,x)\cdot 
\bigl(v^{(1)}(t,x)\part_v+z^{(1)}(t,x)\cdot \nabla_z\bigr)\eta^{j(0)}(t,x) \nn \\
&&+\bigl(v^{(1)}(t,x)\part_v+z^{(1)}(t,x)\cdot \nabla_z\bigr)g^{(0)}(t,x)~,
\eea
where the partial differentials with respect to $v$ and $z$ are taken by considering 
$\mu,\eta$ and $g$ as functions of $(t,x,v,z)$. It is still a linear parabolic PDE.

\subsection{Higher orders and an equivalent simpler formulation}
\label{master-pde-sec}
Although we can proceed to higher orders in the same way and solve $(v^{(i)},z^{(i)})$, there is 
another way with a clearer representation.
Let us define
\be
v^{[i]}(t,x)=\sum_{j=0}^i \ep^j v^{(j)}(t,x), \qquad z^{[i]}(t,x)=\sum_{j=0}^i\ep^j z^{(j)}(t,x)~.
\ee
and the operator
\bea
\call^{[k]}(t,x)&=&\Bigl\{\gamma^l_0\bigl(t,x,v^{[k]}(t,x),z^{[k]}(t,x)\bigr)\part_l+
\frac{1}{2}(\gamma^l\cdot \gamma^m)\bigl(t,x,v^{[k]}(t,x),z^{[k]}(t,x)\bigr)\part_{lm}\Bigr\}\nn \\
\eea
Then, one can easily check that the PDE for $v^{[i]}$ with $(i\geq 1)$ can be expressed as
\bea
\label{master-pde}
&\Bigl(\part_t + \call^{[i-1]}(t,x)\Bigr)v^{[i]}(t,x)+f\bigl(t,x,v^{[i-1]}(t,x),z^{[i-1]}(t,x)\bigr)=0& \\
&v^{[i]}(T,x)=\Phi(x)~
\eea
and
\be
\label{master-z}
z^{[i]}(t,x)=\part_lv^{[i]}(t,x)\gamma^l\bigl(t,x,v^{[i-1]}(t,x),z^{[i-1]}(t,x)\bigr)~.
\ee
It is straightforward to confirm the consistency with the summation of each $(v^{(k)},z^{(k)})$ 
for $(0\leq k\leq i)$ up to the error terms of $o(\ep^i)$, which is due to the additional $\ep$ in front of the 
non-linear terms. Note that, in an arbitrary order, the PDE has a linear parabolic form.
\\

The above formulation clearly shows that the perturbative treatment of 
non-linear effects of the original system allows us to obtain a 
series of linear parabolic PDEs with the same structure.
Solving the PDE for the zero-th order, and then recursively replacing the backward components 
by the solution of the previous expansion order, we can obtain 
an  arbitrary higher order of the approximation.

\section{ Perturbation in Probabilistic Framework for the Generic Coupled Non-linear FBSDEs}
\label{four-step-prob}
We have now seen the perturbation method in the PDE framework
can work even for the fully-coupled non-linear FBSDEs.
In this section, we will provide a corresponding perturbation scheme 
under the probabilistic framework. As we will see, it is nothing more difficult than 
the decoupled case studied in Sec.~\ref{decoupled_approx}, and reduces to the 
standard calculations for the European contingent claims.
As a by-product, applying the asymptotic expansion method explained in Sec.~\ref{asymptotic},
we can also show that it is possible to obtain an analytic expression for the non-linear PDE in the Four Step Scheme
up to the given order of expansion.

\subsection{Generic Formulation}
We try to solve the same FBSDE (\ref{fbsde_org}) treated in the PDE framework. 
Suppose that we have somehow obtained a solution of $(v^{[i-1]}(t,x), z^{[i-1]}(t,x))$.
Then, let us consider the following FBSDE:
\bea
dV^{[i]}_t&=&-f\left(t,X_t^{[i]},v^{[i-1]}(t,X^{[i]}_t),z^{[i-1]}(t,X_t^{[i]})\right)dt+Z^{[i]}_t\cdot dW_t \nn \\
V_T^{[i]}&=&\Phi(X^{[i]}_T) \nn \\
dX^{[i]}_t&=&\gamma_0\left(t,X^{[i]}_t,v^{[i-1]}(t,X^{[i]}_t),z^{[i-1]}(t,X_t^{[i]})\right)dt \nn \\
&&\hspace{20mm}+\gamma\left(t,X^{[i]}_t,v^{[i-1]}(t,X_t^{[i]}),z^{[i-1]}(t,X_t^{[i]})\right)\cdot dW_t \nn\\
X_0^{[i]}&=&x~.
\label{grading-fbsde}
\eea
Here, one can immediately check that the solution of the above FBSDE $(V_t^{[i]}, Z_t^{[i]})$,
as a function of $(t,X_t^{[i]})$, actually satisfies the PDE in the Four Step Scheme given in 
(\ref{master-pde}) and (\ref{master-z}) by setting
\be
v^{[i]}(t,x)=V^{[i]}(t,x),\qquad z^{[i]}(t,x)=Z^{[i]}(t,x).
\ee
Hence the solution of the above FBSDE can be interpreted as the $\ep^i$-th order approximation 
of the original FBSDE in (\ref{fbsde_org}).
Therefore, if we can solve the above FBSDE in probabilistic way, we can proceed to an arbitrarily higher 
order of approximation by simply updating the backward components of the non-linear terms recursively.
We can also say that it is a probabilistic way to solve the non-linear PDE (\ref{fbsde_org})
order by order of $\ep$.

One can check that the above FBSDE is actually decoupled and linear by writing it explicitly as
\bea
&&dV_t^{[i]}= c(t,X_t^{[i]})V_t^{[i]}dt-\ep g(t,X^{[i]}_t,v^{[i-1]}(t,X^{[i]}_t),z^{[i-1]}(t,X^{[i]}_t))dt+Z_t^{[i]}\cdot dW_t \nn \\
&&V_T^{[i]} = \Phi(X_T^{[i]}) \nn \\
&&dX_t^{[i]}= \Bigl(r(t,X_t^{[i]})+\ep \mu\bigl(t,X_t^{[i]},v^{[i-1]}(t,X_t^{[i]}), z^{[i-1]}(t,X_t^{[i]})\bigr)\Bigr)dt\nn \\
&&\hspace{25mm}+\Bigl(\sigma(t,X_t^{[i]})+\ep \eta\bigl(t,X_t^{[i]},v^{[i-1]}(t,X_t^{[i]}),z^{[i-1]}(t,X_t^{[i]})\bigr)\Bigr)\cdot dW_t\nn \\
&&X_0^{[i]}=x~,
\eea
and hence, we can straightforwardly integrate it as
\bea
\label{fbsde-vi}
&&\hspace{-10mm}V_t^{[i]}=E\left[e^{-\int_t^T c(s,X_s^{[i]})ds}\Phi(X_T^{[i]})\right.\nn \\
&&+\left.\left.\ep \int_t^T e^{-\int_t^u c(s,X_s^{[i]})ds}g\bigl(u,X_u^{[i]},v^{[i-1]}(u,X_u^{[i]}),z^{[i-1]}(u,X_u^{[i]})\bigr)
du\right|\calf_t\right] \\
\label{fbsde-zi}
&&\hspace{-10mm}Z_t^{[i]}=E\left[\cald_t\Bigl\{e^{-\int_t^T c(s,X_s^{[i]})ds}\Phi(X_T^{[i]})\right.\nn \\
&&+\left.\left.\ep \int_t^T e^{-\int_t^u c(s,X_s^{[i]})ds}g\bigl(u,X_u^{[i]},v^{[i-1]}(u,X_u^{[i]}),z^{[i-1]}(u,X_u^{[i]})\bigr)
du\Bigr\}\right|\calf_t\right]~.
\eea

The result is equivalent to the pricing of standard European contingent claims, and also has the 
same form appeared in Sec.~\ref{decoupled_approx}.
Thus, we can apply the asymptotic expansion method given in Sec.~\ref{asymptotic} to the 
forward components $X^{[i]}$ in the same way. This will give us the analytical result of 
$(V_t^{[i]},Z_t^{[i]})$ as a function of $(t,X_t^{[i]})$, up to a given order of volatility parameter, say $\del^k$.
Then we can set 
\bea
v^{[i]}(t,x)=V_t^{[i]}(t,x),\qquad z^{[i]}(t,x)=Z_t^{[i]}(t,x)~.
\eea
up to the error terms of $o(\del^k)$, and can move on to the higher order of 
approximations~\footnote{Since we finally put $\del=1$ (and also $\ep=1$), 
the actual order of error terms are of $o((\sigma+\eta)^k)$ in this example.}.

\subsection{Summary of Recursive Procedures}
Here, let us summarize the procedures of our perturbation method.
Firstly, in the zero-th order, the corresponding FBSDE is given by
\bea
dV_t^{[0]}&=&c(t,X_t^{[0]})V_t^{[0]}dt+Z_t^{[0]}\cdot dW_t\\
V_T^{[0]}&=&\Phi(X_T^{[0]})\\
dX_t^{[0]}&=&r(t,X_t^{[0]})dt+\sigma(t,X_t^{[0]})\cdot dW_t \\
X_0^{[0]}&=&x~.
\eea
This can be integrated as
\bea
V_t^{[0]}&=&E\left[\left. e^{-\int_t^T c(s,X_s^{[0]})ds}\Phi(X_T^{[0]})\right|\calf_t\right]\\
Z_t^{[0]}&=&E\left[\left. \cald_t\left(e^{-\int_t^T c(s,X_s^{[0]})ds}\Phi(X_T^{[0]})\right)\right|\calf_t\right]
\eea
which can be solved either exactly, or analytically up to the certain order of volatility 
by the asymptotic expansion method. 
Then we set 
\be
v^{[0]}(t,x)=V^{[0]}(t,x)~,\qquad  z^{[0]}(t,x)=Z^{[0]}(t,x),
\ee
and then put them back in the backward components of (\ref{grading-fbsde}) with $i=1$.
We then obtain (\ref{fbsde-vi}) and (\ref{fbsde-zi}) with $i=1$. We can express 
$V_t^{[1]}$ and $Z_t^{[1]}$ in terms of $t$ and $X_t^{[1]}$
by using the asymptotic expansion method, and use them to define $(v^{[1]}(t,x),z^{[1]}(t,x))$ in turn.
Now, we can move to (\ref{fbsde-vi}) and (\ref{fbsde-zi}) with $i=2$.
We repeat the same procedures to the desired order of approximation.
\\


{\it Remark}: Although we have considered one-dimensional process for $V$, it is 
straight forward to extend the method for higher dimensional cases.
Once we take the basis of $X$ in such a way that the linear drift term $V$ is diagonal,
we can proceed without any difficulty. The mixing from the other components of $V$ always 
appears in the lower order of $\ep$, which keeps the diagonal form of drift term intact
in an arbitrary order.

\section{Conclusion and  Discussions}
In this paper, we have presented a simple perturbation scheme for 
non-linear decoupled as well as coupled FBSDEs. By considering the interested system as 
a decoupled linear FBSDE with non-linear perturbation terms, we succeeded to provide 
the analytic approximation method to an arbitrarily higher order of expansion.
We have shown that the required calculations in each order are equivalent to those 
for standard European contingent claims.
We have applied the method to the two simple models and compared them with
the numerical results directly obtained from the PDE and regression-based Monte Carlo simulation.
Both of the examples clearly demonstrated the strength of our method.
We have also shown that the use of the asymptotic expansion method for forward components allows us to proceed 
to the higher order of perturbation even if the forward components do not have known distributions.

In the last part of the paper, we have studied the perturbative method in the PDE framework based on the 
so-called Four Step Scheme. We have shown that our perturbative treatment 
renders the original non-linear FBSDE into the series of linear parabolic PDEs, which 
are straightforward to handle.
Furthermore, by the equivalence of the two approaches,
we were also able to provide the corresponding  perturbative method in probabilistic framework 
which is explicitly consistent with the Four Step Scheme up to a given order 
of expansion.

The perturbation theory presented in this paper 
may turn out to be crucial to investigate various interesting problems, such as 
those given in the introduction,  which have been preventing analytical treatment so far. 
The application of the new method to the important financial problems is one of our ongoing research topics.

Finally, let us remark on the further extension to the cases including jumps.
Although, in this work,  we have only considered the dynamics driven by Brownian motions, 
the same approximation scheme can also be applied to more generic cases.
Although it will be more difficult to obtain explicit expressions in terms of forward components,
if we choose the specific forward processes with appropriate analytical properties, we should be able to  
proceed in the similar way.  Particularly, the separation of the original system 
into the decoupled linear FBSDE and the non-linear perturbation terms can be done in a
completely parallel fashion.

\appendix
\section{Linear Volatility Term in the Driver}
As we have briefly mentioned in Section~\ref{decoupled_approx}, there are the situations 
where the driver contain the linear term in volatility $Z$~\footnote{We are grateful for an anonymous referee 
to point this out.}. Although it is possible to absorb it
by the measure change, there may be the situation where its direct treatment improves the numerical 
performance than inducing the drift modification in the forward components.
Consider the situation where the backward component follows the following BSDE:
\bea
dV_t=\Bigl(c(t,X_t)V_t+\theta(t,X_t)\cdot Z_t\Bigr)dt-\ep g(t,X_t,V_t,Z_t)dt+Z_t\cdot dW_t~.
\eea
Also in this case, the recursive procedures can be carried out in the same way by simply replacing the factor
\be
\exp\left(-\int_t^s c(u,X_u)du\right) 
\ee
by
\bea
\cale(t,s)=\exp\left(-\int_t^s \Bigl(c(u,X_u)+\frac{1}{2}||\theta(u,X_u)||^2\Bigr)du-\int_t^s \theta(u,X_u)\cdot dW_u\right)~.
\eea
For example, the results of Section~\ref{decoupled_zeroth} can now be expressed as
\bea
V_t^{(0)}&=&E\Bigl[\cale(t,T)\Phi(X_T)\Bigr|\calf_t\Bigr]\\
Z_t^{(0)}&=&E\left[\left.\cald_t\Bigl(\cale(t,T)\Phi(X_T)\Bigr)\right|\calf_t\right]~.
\eea
Higher order and coupled cases can be expressed similarly. In general, it is difficult to say which method performs better,
since both of them have errors of the same order of $\epsilon$ at a given expansion.

\section{Another expansion method for coupled FBSDEs}
In the appendix, we provide another method which is more closely related to that of Sec.\ref{decoupled_approx}
for generic coupled FBSDEs.
We consider the same coupled FBSDE as in Sec.~\ref{four-step-prob}:
\bea
&&\hspace{-12mm}dV_t^{(\ep)}=c(X_t^{(\ep)})V_t^{(\ep)}dt-\ep g(X_t^{(\ep)},V_t^{(\ep)},Z_t^{(\ep)})dt+Z_t^{(\ep)}\cdot dW_t \\
&&\hspace{-12mm}V_T^{(\ep)}=\Phi(X_T^{(\ep)}) \\
&&\hspace{-12mm}dX_t^{(\ep)}=r(X_t^{(\ep)})dt+\ep \mu(X_t^{(\ep)},V_t^{(\ep)},Z_t^{(\ep)})dt
+\left(\sigma(X_t^{(\ep)})+\ep \eta(X_t^{(\ep)},V_t^{(\ep)},Z_t^{(\ep)})\right)\cdot dW_t \\
&&\hspace{-12mm}X_0^{(\ep)}=x ~.
\eea
Here, we absorbed a possible explicit time dependency to $X$.

As in Sec.\ref{decoupled_approx}, we are going to expand the backward as well as 
forward components in terms of $\ep$. Suppose that we have 
\bea
V_t^{(\ep)}&=&V_t^{(0)}+\ep V_t^{(1)}+\ep^2 V_t^{(2)}+ \cdots \\
Z_t^{(\ep)}&=&Z_t^{(0)}+\ep Z_t^{(1)}+\ep^2 Z_t^{(2)}+\cdots \\
X_t^{(\ep)}&=&X_t^{(0)}+\ep X_t^{(1)}+\ep^2 X_t^{(2)}+\cdots ~.
\eea
Now, let us derive each term separately.

\subsection{Zero-th order}
In the zero-th order, the original equation reduces to
a linear and decoupled FBSDE, which will going to serve as the center of expansion:
\bea
dV_t^{(0)}&=&c(X_t^{(0)})V_t^{(0)}dt+Z_t^{(0)}\cdot dW_t \\
V_T^{(0)}&=&\Phi(X_T^{(0)}) \\
dX_t^{(0)}&=&r(X_t^{(0)})dt+\sigma(X_t^{(0)})\cdot dW_t \\
X_0^{(0)}&=&x~.
\eea
$X^{(0)}$ is now the standard Markovian process completely decoupled from the backward components.
We can easily solve the backward components as
\bea
V_t^{(0)}&=&E\left[\left.e^{-\int_t^T c(X_s^{(0)})ds}\Phi(X_T^{(0)})\right|\calf_t\right] \\
Z_t^{(0)}&=&E\left[\left.\cald_t\left(e^{-\int_t^T c(X_s^{(0)})ds}\Phi(X_T^{(0)})\right)\right|\calf_t\right]~.
\eea
As explained in the previous section, we can express $V_t^{(0)}$ and $Z_t^{(0)}$
in terms of $X_t^{(0)}$ by the help of the asymptotic expansion for volatility even if the
process of $X^{(0)}$ does not have known distribution.

\subsection{First order correction}
Now let us consider the dynamics of $V^{(\ep)}-V^{(0)}$ and
$X^{(\ep)}-X^{(0)}$. Following the same arguments in Sec.\ref{decoupled_approx}, 
one can show straightforwardly that
\bea
&&\hspace{-10mm}dV_t^{(1)}=c(X_t^{(0)})V_t^{(1)}dt-g^{(1)}(t)dt+Z_t^{(1)}\cdot dW_t \\
&&\hspace{-10mm}V_T^{(1)}=\Phi^{(1)}(T) \\
&&\hspace{-10mm}dX_t^{(1)}=\nabla_x r(X_t^{(0)})\cdot X_t^{(1)}dt+\mu^{(0)}(t)dt
+\eta^{(1)}(t)\cdot dW_t \\
&&\hspace{-10mm}X_0^{(1)}=0~,
\eea
where we have used shorthand notations:
\bea
\Phi^{(1)}(T)&=&\Phi^{(1)}(X_T^{(0)},X_T^{(1)}) =\nabla_x \Phi(X_T^{(0)})\cdot X_T^{(1)}  \\
g^{(0)}(t)&=&g(X_t^{(0)},V_t^{(0)},Z_t^{(0)}) \\
g^{(1)}(t)&=&g^{(1)}(X_t^{(0)},X_t^{(1)},V_t^{(0)},Z_t^{(0)}) =g^{(0)}(t)-\bigl(\nabla_x c(X_t^{(0)})\cdot X_t^{(1)}\bigr)V_t^{(0)} \\
\mu^{(0)}(t)&=&\mu(X_t^{(0)},V_t^{(0)},Z_t^{(0)}) \\
\eta^{(0)}(t)&=&\eta(X_t^{(0)},V_t^{(0)},Z_t^{(0)}) \\
\eta^{(1)}(t)&=&\eta^{(1)}(X_t^{(0)},X_t^{(1)},V_t^{(0)},Z_t^{(0)})=\eta^{(0)}(t)+\nabla_x \sigma(X_t^{(0)})\cdot X_t^{(1)}.
\eea
Note that, since we have obtained $V_t^{(0)}$ and $Z_t^{(0)}$ as the functions of $X_t^{(0)}$,
the pair $(X^{(0)},X^{(1)})$ consists of a Markovian process, which is indeed decoupled from 
$(V^{(1)},Z^{(1)})$. This means that we have ended up with the decoupled linear FBSDE also for the 
first order correction. Hence, one can easily solve the backward components as
\bea
\label{v1_couple}
&&\hspace{-17mm}V_t^{(1)}=E\left[\left. e^{-\int_t^Tc(X_s^{(0)})ds}\Phi^{(1)}(T)
+\int_t^T e^{-\int_t^u c(X_s^{(0)})ds}g^{(1)}(u)du\right|\calf_t\right] \\
\label{z1_couple}
&&\hspace{-17mm}Z_t^{(1)}=E\left[\left. 
\cald_t\left(e^{-\int_t^Tc(X_s^{(0)})ds}\Phi^{(1)}(T)
+\int_t^T e^{-\int_t^u c(X_s^{(0)})ds}g^{(1)}(u)du\right)\right|\calf_t\right]
\eea
As a result, we have obtained $V_t^{(1)}$as well as $Z_t^{(1)}$ as the functions of 
$X_t^{(0)}$ and $X_t^{(1)}$.

\subsection{Second and Higher order corrections}
We can continue to the higher order corrections in the same way.
By considering $V^{(\ep)}-(V^{(0)}+\ep V^{(1)})$ and 
$X^{(\ep)}-(X^{(0)}+\ep X^{(1)})$, and then extracting the $\ep$-second order terms,
one can show that
\bea
&&\hspace{-17mm}dV_t^{(2)}=c(X_t^{(0)})V_t^{(2)}dt-g^{(2)}(t)dt+Z_t^{(2)}\cdot dW_t \\
&&\hspace{-17mm}V_T^{(2)}=\Phi^{(2)}(T) \\
&&\hspace{-17mm}dX_t^{(2)}=\nabla_x r(X_t^{(0)})\cdot X_t^{(2)}dt+\mu^{(2)}(t)dt+\eta^{(2)}(t)\cdot dW_t\\
&&\hspace{-17mm}X_0^{(2)}=0~.
\eea
Here we have defined
\bea
\Phi^{(2)}(T)&=&\Phi^{(2)}(X_T^{(0)},X_T^{(1)},X_T^{(2)}) \\
&=&\nabla_x\Phi(X_T^{(0)})\cdot X_T^{(2)}+\frac{1}{2}\part_{ij}\Phi(X_T^{(0)})X_T^{i(1)}X_T^{j(1)} \\
&&\hspace{-25mm}g^{(2)}(t)=g^{(2)}(X_t^{(0)},X_t^{(1)},X_t^{(2)},V_t^{(0)},V_t^{(1)},Z_t^{(0)},Z_t^{(1)}) \\
&&\hspace{-23mm}=\left\{\nabla_x g^{(0)}(t)\cdot X_t^{(1)}+\frac{\part}{\part v}
g^{(0)}(t)V_t^{(1)}+\nabla_z g^{(0)}(t)\cdot Z_t^{(1)}\right\}\nn \\
&&\hspace{-23mm}-\Bigl(\nabla_x c(X_t^{(0)})\cdot X_t^{(2)}+\frac{1}{2}\part_{ij}c(X_t^{(0)})X_t^{i(1)}X_t^{j(1)}\Bigr)V_t^{(0)}
-\left(\nabla_x c(X_t^{(0)})\cdot X_t^{(1)}\right)V_t^{(1)}. \nn \\
\eea
And also for the process $X^{(2)}$, we have defined 
\bea
\mu^{(2)}(t)&=&\mu^{(2)}(X_t^{(0)},X_t^{(1)},X_t^{(2)},V_t^{(0)},V_t^{(1)},Z_t^{(0)},Z_t^{(1)})\\
&=&\Bigl\{\nabla_x \mu^{(0)}(t)\cdot X_t^{(1)}+
\frac{\part}{\part v}\mu^{(0)}(t) V_t^{(1)}+\nabla_z \mu^{(0)}(t)\cdot Z_t^{(1)}\Bigr\} \\
&&+\frac{1}{2}\part_{ij}r(X_t^{(0)})X_t^{i(1)}X_t^{j(1)}
\eea
and
\bea
\eta^{(2)}(t)&=&\eta^{(2)}(X_t^{(0)},X_t^{(1)},X_t^{(2)},V_t^{(0)},V_t^{(1)},Z_t^{(0)},Z_t^{(1)})\\
&=&\nabla_x \sigma(X_t^{(0)})\cdot X_t^{(2)}+
\frac{1}{2}\part_{ij}\sigma(X_t^{(0)})X_t^{i(1)}X_t^{j(1)} \nn \\
&+&\Bigl\{\nabla_x\eta^{(0)}(t)X_t^{(1)}+\frac{\part}{\part v}\eta^{(0)}(t)V_t^{(1)}+
\nabla_z\eta^{(0)}(t)\cdot Z_t^{(1)}\Bigr\}~.
\eea

One can check that the pair $(X_t^{(0)},X_t^{(1)},X_t^{(2)})$ consists of a Markovian process
and it is decoupled from the backward components. Hence, once again, we have obtained
the decoupled linear FBSDE, which is solvable as before:
\bea
V_t^{(2)}&=&E\left[\left.e^{-\int_t^T c(X_s^{(0)})ds}\Phi^{(2)}(T)+\int_t^Te^{-\int_t^u c(X_s^{(0)})ds}
g^{(2)}(u)du
\right|\calf_t\right] \\
Z_t^{(2)}&=&E\left[\left.\cald_t\Bigl\{ e^{-\int_t^T c(X_s^{(0)})ds}\Phi^{(2)}(T)+\int_t^Te^{-\int_t^u c(X_s^{(0)})ds}g^{(2)}(u)du\Bigr\}
\right|\calf_t\right]~. 
\eea
In completely the same way, we can proceed to an arbitrarily higher order correction.
In each expansion order $\ep^i$, the set $(X_t^{(0)}, \cdots, X_t^{(i)})$ follows
Markovian process decoupled from the backward components, and 
also the FBSDE continues to be linear thank to the $\ep$ in front of the non-linear terms.
\\
\\
{\it Remark}: For simplicity of the presentation, we have used the common 
perturbation parameter $\ep$ both for the non-linearity in backward components
as well as the feedback effects in the forward components.
However, as one can easily expect, it is also possible to introduce multiple perturbation
parameters. 

\subsection{Consistency to the result of Sec.~\ref{four-step-prob}}
For completeness, let us check the consistency to the result of 
Sec.~\ref{four-step-prob}.
In the zero-th order, the corresponding 
FBSDEs are exactly equal.
In the first order, we have obtained
\bea
V_t^{[1]}&=&E\left[e^{-\int_t^T c(s,X_s^{[1]})ds}\Phi(X_T^{[1]})\right.\nn \\
&&\quad \left.\left.+\ep \int_t^T e^{-\int_t^u c(s,X_s^{[1]})ds}g\bigl(u,X_u^{[1]},v^{(0)}(u,X_u^{[1]}),z^{(0)}(u,X_u^{[1]})\bigr)
du\right|\calf_t\right]
\label{1st_order_n}
\eea
and $Z_t^{[1]}$ as its Malliavin derivative in Sec.~\ref{four-step-prob}.
Now, in order to compare it to the current method, let us consider
\be
V_t^{[1]}-V_t^{(0)}
\ee
and extract $\ep$ first order terms by expanding $X_t^{[1]}=X_t^{(0)}+\ep X_t^{(1)}$.
Since the second term of (\ref{1st_order_n}) has already $\ep$ in the front,
we simply get a contribution from there as
\be
E\left[\left.\int_t^T e^{-\int_t^u c(s,X_s^{(0)})ds}g\bigl(u,X_u^{(0)},v^{(0)}(u,X_u^{(0)}),z^{(0)}(u,X_u^{(0)})\bigr)du
\right|\calf_t\right].
\ee
From the first term, we have to expand $X^{[1]}$ in the terminal payoff and also in the discount:
\bea
E\left[\left.e^{-\int_t^T c(s,X_s^{(0)})ds}\nabla_x \Phi(X_T^{(0)})\cdot X_T^{(1)}-
\int_t^T e^{-\int_t^u c(s,X_s^{(0)})ds}V_u^{(0)}\bigl(\nabla_x c(u,X_u^{(0)})\cdot X_u^{(1)}\bigr)du\right|\calf_t\right]\nn
\eea
where we have used the Gateaux derivative~\footnote{See, for examples, \cite{Duffie} for similar calculation.}.
Second term can be interpreted that there is a change in value by
\be
-\bigl(\nabla_x c(u,X_u^{(0)})\cdot X_u^{(1)}\bigr)V_u^{(0)}
\ee
at each point of time $u$, which is summed and discounted back to the current time.
By summing the above two terms, one can easily confirm its equivalence to the result of the 
previous section. Applying Malliavin derivative automatically tells the consistency of volatility terms.
Using the same arguments, one can check the consistency between the 
current method and that of Sec.~\ref{four-step-prob}. Although we have solved
$V^{(i)}$ separately, the sum $\sum^i_{k=0}\ep^k V^{(k)}$ can be shown equivalent to $V^{[i]}$
up to the error terms $o(\ep^i)$.



\begin{thebibliography}{99}

\bibitem{Bismut}
Bismut, J.M. 1973, "Conjugate Convex Functions in Optimal Stochastic Control," J. Political
Econ., 3, 637-654.

\bibitem{Duffie_Epstein}
Duffie, D. and Epstein, L., 1992, "Stochastic Differential Utility,"
Econometrica 60 353-394.

\bibitem{Duffie}
Duffie, D., Huang, M., 1996, 
"Swap Rates and Credit Quality," Journal of Finance,
Vol. 51, No. 3, 921.

\bibitem{ElKaroui}
El Karoui, N., Peng, S.G., and Quenez, M.C., 1997, "Backward stochastic 
differential equations in finance," Math. Finance $\bold{7}$ 1-71.


\bibitem{asymmetric_collateral}
Fujii, M., Takahashi, A., 2010, "Derivative pricing under Asymmetric and 
Imperfect Collateralization and CVA,"
CARF Working paper series F-240, available at
http://ssrn.com/abstract=1731763.


\bibitem{Gobet}
Gobet, E., Lemor, J., Warin, X., 2005, "A Regression-based Monte Carlo Method
to solve Backward Stochastic Differential Equations, "
The Annals of Applied Probability, 15, No.3, 2172-2202.

\bibitem{KT}
Kunitomo, N. and Takahashi, A. 2003,
"On Validity of the Asymptotic Expansion Approach
 in Contingent Claim Analysis,"
Annals of Applied Probability, 13, No.3, 914-952.

\bibitem{fourstep} 
Ma, J., Protter, P., and Yong, J., 1994, "Solving forward-backward stochastic differential equations explicitly",
Prob.\& Related Fields, 98.

\bibitem{Ma}
Ma, J., and Yong, J., 2000, "Forward-Backward Stochastic Differential Equations and their
Applications," Springer.

\bibitem{Peng1}
Pardoux, E., and S. Peng, 1990, "Adapted Solution of a Backward Stochastic Differential Equation,"
Systems Control Lett., 14, 55-61.


\bibitem{STT}
Shiraya, K., Takahashi, A., and Toda, M, 2009,
"Pricing Barrier and Average Options under Stochastic Volatility Environment," 
CARF Working Paper F-242, available at http://www.carf.e.u-tokyo.ac.jp/workingpaper/,
forthcoming in Journal of Computational Finance.

\bibitem{T}
Takahashi, A. 1999,
"An Asymptotic Expansion Approach to Pricing Contingent Claims," 
Asia-Pacific Financial Markets, 6, 115-151.

\bibitem{asymptotic3}
Takahashi, A., Takehara, K., and Toda, M., 2011, "A General Computation Scheme for a High-Order Asymptotic Expansion Method,"
CARF Working Paper F-242, available at http://www.carf.e.u-tokyo.ac.jp/workingpaper/.

\bibitem{TY}
Takahashi, A. and Yoshida, N. 2004, 
``An Asymptotic Expansion Scheme for Optimal Investment Problems,''
Statistical Inference for Stochastic Processes, 7, No.2, 153-188.


\bibitem{zhou}
Yong, J., and Zhou., X., 1999, "Stochastic Control," Springer


\end{thebibliography}
\end{document}